\documentclass{article}

\usepackage{a4}

\usepackage[vcentermath]{youngtab}

\newcommand{\Y}{\yng}

\newcommand{\diag}{{\rm diag}}
\newcommand{\tr}{{\rm tr}}
\newcommand{\ol}{\overline}

\newcommand{\wt}{\widetilde}

\usepackage{amsmath}
\usepackage{amsfonts}

\newcommand{\RR}{\mathbb{R}}
\newcommand{\CC}{\mathbb{C}}
\newcommand{\ZZ}{\mathbb{Z}}

\newcommand{\cSU}{{{\cal S}_U}}
\newcommand{\cST}{{{\cal S}_T}}
\newcommand{\cS}{{\cal S}}

\def \be {\begin{equation}}
\def \ee {\end{equation}}
\def \bea {\begin{eqnarray}}
\def \eea {\end{eqnarray}}

\title{A Monopole Index for ${\cal N}=4$ Chern-Simons Theories}
\author{Y.~Imamura and S.~Yokoyama}
\begin{document}

\begin{titlepage}
\title{\hfill\parbox{4cm}
       {\normalsize UT-09-08\\July 2009}\\
       \vspace{2cm}
A Monopole Index for ${\cal N}=4$ Chern-Simons Theories
       \vspace{2cm}}
\author{
Yosuke Imamura\thanks{E-mail: \tt imamura@hep-th.phys.s.u-tokyo.ac.jp}
and Shuichi Yokoyama\thanks{E-mail: \tt yokoyama@hep-th.phys.s.u-tokyo.ac.jp}
\\[30pt]
{\it Department of Physics, University of Tokyo,}\\
{\it Hongo 7-3-1, Bunkyo-ku, Tokyo 113-0033, Japan}
}
\date{}

\maketitle
\thispagestyle{empty}

\vspace{0cm}

\begin{abstract}
\normalsize
We compute a certain index for an ${\cal N}=4$ Chern-Simons theory
with gauge group $U(N)^r$ in the large $N$ limit
with taking account of monopole contribution,
and compare it to the corresponding
multi-particle index for M-theory in
the dual geometry AdS$_4\times X_7$.
The internal space $X_7$ has non-trivial two-cycles,
and M2-branes wrapped on them contribute to the
multi-particle index.
We establish one-to-one map between $r-1$ independent magnetic charges
on the gauge theory side and the same number of charges
on the gravity side:
the M-momentum and $r-2$ ($=b_2(X_7)$) wrapping numbers.
With a certain assumption for the wrapped M2-brane contribution,
we confirm the agreement of the indices
for many sectors specified by the $r-1$ charges
by using analytic and numerical methods.
\end{abstract}

\end{titlepage}

\section{Introduction}\label{intro.sec}
Monopole operators play important roles in
gauge theories for M2-branes.
Taking account of them is essential to obtain correct background geometries
of M2-branes as moduli spaces of the gauge theories.
Let us consider the ${\cal N}=8$ supersymmetric Yang-Mills theory
realized on a stack of $N$ D2-branes as a simple example.
The diagonal $U(1)$ subgroup of the $U(N)$ gauge group
does not couple to any fields, and we can define
the dual photon field $a$ by
\begin{equation}
da=\frac{1}{g_3^2}*\tr F,
\end{equation}
where $g_3$ is the gauge coupling.
This scalar field describes the collective motions
of the M2-branes, and is identified with ``the eleventh direction''
or ``the M-direction.''
By definition, $a$ is the canonical conjugate of $\tr F$,
and the operator $e^{ima}$ changes the flux $(2\pi)^{-1}\int\tr F$ by $m$.
Namely, $e^{ima}$ is a monopole operator carrying magnetic charge $m$.
In this sense,
the M-direction emerges by taking account of monopole operators.

The ABJM model\cite{Aharony:2008ug}, $U(N)\times U(N)$
Chern-Simons matter system with ${\cal N}=6$ supersymmetry,
was proposed as a theory describing M2-branes in $\RR^8/\ZZ_k$.
Monopole operators in ABJM model\cite{Berenstein:2008dc,Hosomichi:2008ip,Klebanov:2008vq,Kim:2009wb,Benna:2009xd,Berenstein:2009sa,Kim:2009ia}
also play a similar role.
The action of ABJM model includes Chern-Simons terms
\begin{equation}
S_{\rm CS}=\frac{k}{4\pi}\int\tr\left[
A_1dA_1+\frac{2}{3}A_1^3
-A_2dA_2-\frac{2}{3}A_2^3\right].
\end{equation}
The component of gauge fields corresponding to the diagonal $U(1)$ subgroup
of $U(N)\times U(N)$ appears in the action only through these
Chern-Simons terms, and its equation of motion gives
the Gauss-law constraint
\begin{equation}
\tr F_1-\tr F_2=0.
\label{abjmgauss}
\end{equation}
Therefore, we can define one gauge invariant magnetic charge again by
\begin{equation}
m=\frac{1}{2\pi}\oint\tr F_1=\frac{1}{2\pi}\oint\tr F_2.
\end{equation}
Just as in the case of the D2-brane theory,
this magnetic charge is identified with the Kaluza-Klein momentum along
the M-direction (M-momentum).
The dual photon field $a$
can be defined by solving
(\ref{abjmgauss}) as
\begin{equation}
da=\tr A_1-\tr A_2,
\label{abjmadef}
\end{equation}
and monopole operators in the form $e^{ima}$ correspond to
Kaluza-Klein modes with the M-momentum proportional to $m$.
(More precisely, the operator $e^{ima}$ is not gauge invariant in the ABJM model.
We can construct gauge invariant operators by combining $e^{ima}$ and matter fields,
and such gauge invariant operators correspond to Kaluza-Klein modes.)
See also \cite{Lambert:2008et,Distler:2008mk}
for similar analysis for BLG model\cite{Bagger:2006sk,Bagger:2007jr,Bagger:2007vi,Gustavsson:2007vu,Gustavsson:2008dy}.

When $k=1$ or $2$, the supersymmetry of the ABJM model is
expected to be enhanced from ${\cal N}=6$ to ${\cal N}=8$.
Equivalently, the R-symmetry is enhanced from $SO(6)$ to $SO(8)$.
Because the $SO(8)$ mixes the M-direction with other directions,
the analysis of monopole operators is indispensable to show the existence of
the ${\cal N}=8$ supersymmetry\cite{Gustavsson:2009pm,Kwon:2009ar}.

Even in more general quiver Chern-Simons theories,
the emergence of the M-direction is explained in the same way\cite{Jafferis:2008qz,Hanany:2008cd,Ueda:2008hx,Imamura:2008qs}.
In quiver-type theories,
we always have $U(1)$ diagonal gauge field which couples no
matter fields in the theory and its dual scalar field plays the role of
the coordinate of the M-direction.
In general, however, we can define more than one magnetic charges,
and only one of them accounts for the emergence of the M-momentum.
For more concrete analysis, let us consider a $U(N)^r$ quiver Chern-Simons theory
with the Chern-Simons terms
\begin{equation}
S_{\rm CS}=\sum_{a=1}^r\frac{k_a}{4\pi}\int\left(A_adA_a+\frac{2}{3}A_a^3\right).
\label{scs}
\end{equation}
We assume that the levels $k_a$ satisfy
\begin{equation}
\sum_{a=1}^rk_a=0.
\label{totalk0}
\end{equation}
In this theory, 
we can define $r$ gauge invariant magnetic charges by
\begin{equation}
m_a=\frac{1}{2\pi}\oint\tr F_a,\quad
a=1,\ldots,r.
\label{madef}
\end{equation}
But there is one constraint imposed on these magnetic charges.
Due to the assumption (\ref{totalk0}),
the diagonal $U(1)$ gauge field becomes
a Lagrange multiplier, and
its equation of motion gives the constraint
\begin{equation}
\sum_{a=1}^r k_a\tr F_a=0.
\label{gausslaw}
\end{equation}
By integrating this over a $2$-cycle,
we obtain
\begin{equation}
\sum_{a=1}^r k_am_a=0.
\label{kmzero}
\end{equation}
This relation decreases the number of independent magnetic charges by one.
Hence we have $r-1$ independent magnetic charges.
One of them should be identified with the M-momentum,
but we still have extra $r-2$ charges.
What do these charges represent in the dual geometry AdS$_4\times X_7$?

In \cite{Imamura:2008ji,Imamura:2009ur}, it is proposed that
monopole operators carrying
these extra magnetic charges correspond to M2-branes
wrapped on non-trivial two-cycles in the internal space $X_7$.
In \cite{Imamura:2008ji},
it is pointed out that
for ${\cal N}=4$ Chern-Simons theories\cite{Gaiotto:2008sd,Fuji:2008yj,Hosomichi:2008jd,Imamura:2008dt},
the two-cycle Betti number $b_2(X_7)$ and $r-2$,
the number of independent magnetic charges
subtracted by one, agree.
In \cite{Imamura:2009ur}, for Abelian ${\cal N}=4$ theories,
the agreement of the spectrum of monopole operators and that of wrapped
M2-branes are partially confirmed.

The purpose of this paper is to investigate the correspondence between
wrapped M2-branes and monopole operators in more detail.
The analysis in \cite{Imamura:2009ur} is carried out with the gauge group
$U(1)^r$, and by this reason, the perfect agreement of spectrum cannot be
expected.
In this paper, we consider an ${\cal N}=4$ Chern-Simons theory
with gauge group $U(N)^r$, and confirm a certain index for the gauge theory
in the large $N$
limit including monopole contribution agrees with the corresponding multi-particle index on the
gravity side with taking account of contribution of wrapped M2-branes.

Analysis of indices has been done for ABJM model in
\cite{Bhattacharya:2008bja}.
The ABJM model has $SU(4)$ R-symmetry and $U(1)$ flavor symmetry.
Let $(h_1,h_2,h_3)$ be the $SU(4)_R$ weight vector
and $h_4$ be the $U(1)$ charge.
$h_4$ is identified with
the M-momentum on the gravity side.
See Table \ref{coords} for concrete definition of the charges $h_m$.
The superconformal indices investigated in \cite{Bhattacharya:2008bja}
are defined by%
\footnote{We replace the variable $x$ commonly used
in the literature by $x^2$ to avoid fractional power.
We also use different numbering for the chemical potentials $y_i$
from references for convenience.}
\begin{equation}
I(x,y_1,y_2)=\tr\left[(-1)^Fe^{-\beta'\{Q,S\}} x^{2(\Delta+j_3)}y_1^{h_1}y_2^{h_2}\right]
\label{iabjmm0}
\end{equation}
where $Q$ is the component of supercharge with R-charge
$(h_1,h_2,h_3)=(0,0,1)$, and $S$ is its Hermitian conjugate.
On the gauge theory side,
the trace in (\ref{iabjmm0})
is regarded as the summation over gauge invariant operators,
and we denote the index by $I^{\rm gauge}$.
On the gravity side, we can define two indices by (\ref{iabjmm0}).
The single-particle index $I^{\rm sp}$ is defined
by taking the trace over all single-particle states,
while the multi-particle index $I^{\rm mp}$ is also defined by
the same equation by summing up all the multi-particle states
including single- and no-particle states.
In general, a single particle index and the corresponding multi-particle
index are related by
\begin{equation}
I^{\rm mp}(\cdot)
=\exp\sum_{n=1}^\infty\frac{1}{n}I^{\rm sp}(\cdot^n),
\label{impisp}
\end{equation}
where ``$(\cdot)$'' represents the sequence of the arguments of the index,
and ``$(\cdot^n)$'' on the right hand side is the sequence with every argument
replaced by its $n$-th power.

The indices defined by (\ref{iabjmm0}) are
independent of $\beta'$,
and only operators (states) saturating the BPS bound
\begin{equation}
\{Q,S\}=\Delta-j_3-h_3\geq0
\end{equation}
contribute to the indices.
Because the computation of gauge theory index
in \cite{Bhattacharya:2008bja}
is performed
in the large $k$-limit,
all the monopole contribution decouples.
On the gravity side, this corresponds to the decoupling of
Kaluza-Klein modes with non-vanishing M-momentum $h_4$.
The gauge theory index obtained in \cite{Bhattacharya:2008bja}
is
\begin{equation}
I^{\rm gauge}(x,y_1,y_2)=\prod_{n=1}^\infty\frac{(1-x^{4n})^2}{
(1-\frac{x^{2n}}{y_1^n})
(1-\frac{x^{2n}}{y_2^n})
(1-x^{2n} y_1^n)
(1-x^{2n} y_2^n)}.
\label{abjmgaugenom}
\end{equation}
The corresponding graviton index is obtained
by a projection of the graviton index for AdS$_4\times{\bf S}^7$.
The graviton index for AdS$_4\times{\bf S}^7$\cite{Bhattacharya:2008zy} is
\begin{equation}
I^{\rm grav}(x,y_1,y_2,y_3)
=\tr\left[ (-)^F e^{-\beta'\{Q,S\}}
x^{2(\Delta+j_3)}
y_1^{h_1}
y_2^{h_2}
y_3^{h_4}\right]
=\frac{(\mbox{numerator})}{(\mbox{denominator})},
\label{igrav0}
\end{equation}
where the numerator and the denominator are given by
\begin{eqnarray}
(\mbox{numerator})
&=&
\sqrt{y_1y_2y_3}(1+y_1y_2+y_2y_3+y_3y_1)x
\nonumber\\&&
-\sqrt{y_1y_2y_3}(y_1+y_2+y_3+y_1y_2y_3)x^7
\nonumber\\&&
+(y_1y_2+y_2y_3+y_3y_1+y_1y_2y_3(y_1+y_2+y_3))(x^6-x^2),\\
(\mbox{denominator})
&=&(1-x^4)
(\sqrt{y_3}-x\sqrt{y_1y_2})
(\sqrt{y_1}-x\sqrt{y_2y_3})
\nonumber\\&&\times
(\sqrt{y_2}-x\sqrt{y_3y_1})
(\sqrt{y_1y_2y_3}-x).
\label{adsisp}
\end{eqnarray}
The single-particle index for the orbifold ${\bf S}^7/\ZZ_k$
in the large $k$ limit
is obtained from (\ref{igrav0}) by picking up $y_3$ independent terms
as
\begin{equation}
I^{\rm sp}(x,y_1,y_2)=\frac{x^2}{y_1-x^2}+\frac{1}{1-x^2y_1}+\frac{x^2}{y_2-x^2}
+\frac{1}{1-x^2y_2}-\frac{2}{1-x^4}.
\label{ispnomon}
\end{equation}
It is easy to see the perfect agreement between
(\ref{abjmgaugenom}) and the multi-particle index
$I^{\rm mp}(x,y_1,y_2)$ obtained from (\ref{ispnomon}) by
the relation (\ref{impisp}).

A gauge theory index for ABJM model including the monopole contribution
is computed in \cite{Kim:2009wb}.
The index is defined by
\begin{equation}
I^{\rm gauge}(x,y_1,y_2,y_3)=\tr\left[
(-1)^F e^{-\beta'\{Q.S\}}x^{2(\Delta+j_3)}y_1^{h_1}y_2^{h_2}
y_3^{h_4}\right],
\end{equation}
where $y_3$ is introduced as the chemical potential for the charge
$h_4$, which is related to the monopole charge $m$ by $h_4=km$.
In \cite{Kim:2009wb} it is confirmed that
this agrees with the multi-particle index
(\ref{impisp}) with $I^{\rm sp}$ being replaced by $\ZZ_k$ projection
of (\ref{igrav0}).

For ${\cal N}=4$ theories,
a superconformal index without monopole contribution is
computed in \cite{Choi:2008za}.
They consider an ${\cal N}=4$ theory obtained as $\ZZ_M$ orbifold of ABJM model,
which includes $M$ untwisted and $M$ twisted hypermultiplets,
and compute the index
\begin{equation}
I^{\rm gauge}(x,y_1)=\tr\left[(-1)^Fe^{-\beta'\{Q,S\}}x^{2(\Delta+j_3)}y_1^{h_1}\right]
\label{n4sfindex}
\end{equation}
on the gauge theory side.
The result suggests that
the corresponding single-particle index should be
\begin{eqnarray}
I^{\rm sp}(x,y_1)
&=&\frac{1}{1-x^2y_1}+\frac{1}{1-x^2/y_1}-\frac{2}{1-x^4}
+\frac{2x^{2M}}{1-x^{2M}}
\nonumber\\
&&+(M-1)\left(\frac{1}{1-x^2y_1}+\frac{1}{1-x^2/y_1}-\frac{2}{1-x^4}\right).
\label{n4isp}
\end{eqnarray}
An interesting feature of this result is that
this index consists of two parts of different origins.
The first line in (\ref{n4isp})
is obtained
from (\ref{ispnomon}) by the projection
which leaves only terms invariant under the
$\ZZ_M$ rotation $y_2\rightarrow e^{2\pi i/M}y_2$.
Thus, the first line is regarded as the bulk contribution.
On the other hand,
the second line is interpreted as the contribution of twisted sectors.
Indeed, there are two $A_{M-1}$-type singular loci in the internal space
$X_7$, and we expect the existence of an $SU(M)$ vector multiplet on
each of these singular loci.
Let $G_\cS=SU(M)\times SU(M)$ be the gauge group realized on the loci,
and $H_\cS=U(1)^{2(M-1)}$ its Cartan subgroup.
In \cite{Choi:2008za} the twisted sectors
are identified with the contribution of
the $H_\cS$ vector multiplets,
which arise as
supergravity modes
localized at the singular loci.

In this paper
we consider an ${\cal N}=4$ Chern-Simons theory with
general numbers of untwisted and twisted hypermultiplets.
Let $p$ and $q$ be the numbers of two kinds of hypermultiplets.
In this case, as we will explain later,
the internal space $X_7$ includes an $A_{p-1}$-type singular locus
and an $A_{q-1}$-type singular locus.
On these loci gauge group $G_\cS=SU(p)\times SU(q)$ is realized.
We compute an index $I^{\rm gauge}$ similar to (\ref{n4sfindex})
with taking account of the monopole contribution,
and compare it to the corresponding multi-particle index
$I^{\rm mp}$ on the gravity side.
We find that they agree if we take account of
the full $G_\cS$ vector multiplets,
which include not only the Cartan part but also $H_\cS$-charged particles.
The $H_\cS$ charges are represented as a vector in the
$G_\cS$ root lattice,
and we can regard them as wrapping numbers of M2-branes
on the vanishing two-cycles at the singularities.
By comparing the indices,
we establish the relation between
$r-1$ independent magnetic charges and the same number of
charges on the gravity side:
the M-momentum and $r-2$ wrapping numbers.
The $H_\cS$ charges couple to the Wilson lines
associated with the non-trivial fundamental groups of
the singular loci,
and this coupling shifts the Kaluza-Klein momenta
along the cycles.
Such a shift of momenta is correctly reproduced
on the gauge theory side
as a selection rule for charges associated with
the global symmetries $U(1)\times U(1)'$, which are defined in \S\ref{n4cs}.
For $U(1)^r$ case, it was confirmed in \cite{Imamura:2009ur}
by investigating charges associated with $U(1)\times U(1)'$
of gauge-invariant chiral monopole operators.
We show that it is also the case for $U(N)^r$
from the analysis of a monopole index in \S\ref{gauge.sec}.

This paper is organized as follows.
In \S\ref{n4cs}, we briefly review the field contents,
global symmetries, and the moduli space of the ${\cal N}=4$ Chern-Simons theory.
We also comment on the relation between the Wilson lines in the singular loci
and the fivebrane linking numbers.
In \S\ref{grav.sec} we derive a general expression for the
multi-particle index from the known result of the
graviton index for AdS$_4\times{\bf S}^7$ and
a certain assumption for the index for a
vector multiplet localized on AdS$_4\times{\bf S}^3$.
In \S\ref{gauge.sec} we investigate the gauge theory index
including the monopole contribution.
We derive the selection rules for the $U(1)\times U(1)'$ charges
and confirm that they agree with what is expected from the
relation between the Wilson lines and the fivebrane linking numbers.
We compare the gauge theory index and multi-particle index in
\S\ref{compare.sec}
for many sectors specified by the $r-1$ charges,
and confirm that they agree with each other
by using analytic and numerical methods.
We summarize the results in \S\ref{conc.sec}.

\section{${\cal N}=4$ Chern-Simons theories}\label{n4cs}
\subsection{Action and symmetries}
We consider an ${\cal N}=4$ Chern-Simons theory with unitary-type gauge group $U(N)^r$.
Such a Chern-Simons theory is described by a circular quiver diagram,
in which vertices and edges represent vector multiplets $V_a$ and
hypermultiplets $H_I$, respectively.
We use index $a$ for vertices and $I$ for edges.
In the terminology of ${\cal N}=2$ superspace formalism,
${\cal N}=4$ supermultiplets are represented as pairs of superfields
as $V_a=(v_a,\Phi_a)$ and $H_I=(Q_I,\wt Q_I)$.
We define component fields for these superfields as
\begin{equation}
Q_I=(q_I,\psi_I),\quad
\wt Q_I=(\wt q_I,\wt\psi_I),\quad
v_a=(A_\mu^a,\lambda_a),\quad
\Phi_a=(\phi_a,\chi_a).
\end{equation}
The action is
\begin{eqnarray}
S
&=&\sum_{I=1}^r\int d^3xd^4\theta\tr
(Q_I^\dagger e^{2v_{L(I)}}Q_I e^{-2v_{R(I)}}
+\wt Q_I e^{2v_{L(I)}}\wt Q_I^\dagger e^{-2v_{R(I)}})
\nonumber\\&&
-\sum_{I=1}^r\left(\int d^3xd^2\theta\sqrt2\tr(
\wt Q_I\Phi_{L(I)}Q_I
-\wt Q_IQ_I\Phi_{R(I)})+\mbox{c.c.}\right)
\nonumber\\
&&
+\sum_{a=1}^r
\frac{k_a}{2}\tr\left[\frac{1}{2\pi}\int d^3xd^4\theta\int_0^1dt
(v_a\ol D(e^{-2tv_a}De^{2tv_a}))
-\left(\int d^3xd^2\theta\Phi_a^2+\mbox{c.c.}\right)
\right],
\label{action}
\end{eqnarray}
where $L(I)$ and $R(I)$ represent
the vertices at the left and the right ends of an edge $I$,
respectively.
Similarly, we define $L(a)$ and $R(a)$ for
the edges on the left and the right side of a
vertex $a$.
The first line in (\ref{action}) includes
the kinetic terms of the hypermultiplets,
and the second line is the standard superpotential coupling between
vector and hypermultiplets.
The third line is the supersymmetric completion of the Chern-Simons terms
(\ref{scs}).

$k_a$ in (\ref{action}) are the Chern-Simons couplings.
The gauge invariance of the action requires them to be integers.
Furthermore, the existence of ${\cal N}=4$ supersymmetry requires
them to be given by
\be
k_a = k ( s_{L(a)} - s_{R(a)} ),\quad
k\in\ZZ,\quad
s_I=0,1,
\label{kkss}
\ee
where $s_I$ are integers assigned to edges in the quiver diagram,
and they take only two values $0$ and $1$.
Corresponding to these two values,
the hypermultiplets fall into two groups,
untwisted and twisted hypermultiplets.
If $s_I=0$ ($s_I=1$) the hypermultiplet is called untwisted (twisted) hypermultiplet.
When we want to distinguish these two kinds of hypermultiplets,
we use index $i$ for untwisted hypermultiplets,
and $i'$ for twisted ones.
Let $p$ and $q$ be the numbers
of untwisted, and twisted hyper-multiplets,
respectively.
Because the quiver diagram is circular,
the total number of hypermultiplets and the number of vector multiplets
are the same;
\be
r= p+q.
\ee

The R-symmetry of this ${\cal N}=4$ theory is
\be
Spin(4)_R = SU(2) \times SU(2)'.
\ee
There also exist flavor symmetries
\be
U(1) \times U(1)'.
\ee
The component fields in untwisted hypermultiplets $H_i$ and
those of twisted hypermultiplets $H_{i'}$ are
transformed in different ways under the global symmetries.
See Table \ref{globalsym}.
\begin{table}[htb]
\caption{The global symmetries}
\label{globalsym}
\begin{center}
\begin{tabular}{ccccc}
\hline
\hline
& $SU(2)$ & $SU(2)'$ & $U(1)$ & $U(1)'$ \\
\hline
$(q_i,\wt q^\dagger_i)$ & ${\bf2}$ & ${\bf1}$ & $+1$ & $0$ \\
$(\psi_i,\wt\psi_i^\dagger)$ & ${\bf1}$ & ${\bf2}$ & $+1$ & $0$ \\
$(q_{i'},\wt q_{i'}^\dagger)$ & ${\bf1}$ & ${\bf2}$ & $0$ & $+1$ \\
$(\psi_{i'},\wt\psi_{i'}^\dagger)$ & ${\bf2}$ & ${\bf1}$ & $0$ & $+1$\\
\hline
\end{tabular}
\end{center}
\end{table}

The M2-brane background corresponding to this theory is
obtained as the Higgs branch of moduli spaces\cite{Imamura:2008nn}.
(See also \cite{Benna:2008zy,Terashima:2008ba}.)
To obtain the background for a single M2-brane,
let us consider the Abelian case with the gauge group $U(1)^r$.
With the terminology of ${\cal N}=2$ supersymmetry,
the moduli space is obtained by dividing the
solution of F-term conditions by the complexified gauge symmetry.

From the superpotential terms in the action (\ref{action}),
the F-term conditions of $Q_I$ and $\wt Q_I$ give
\begin{equation}
\phi_{L(I)}=\phi_{R(I)}.
\label{phpjh}
\end{equation}
(We assume that $q_I$, $\wt q_I\neq0$ for the Higgs branch.)
This means that all $\phi_a$ take the same value.
We denote it by $\phi$.
The F-term condition for $\Phi_a$ is
\begin{equation}
q_{L(a)}\wt q_{L(a)}-ks_{L(a)}\phi
=q_{R(a)}\wt q_{R(a)}-ks_{R(a)}\phi.
\label{qqksp}
\end{equation}
This means that
$q_I\wt q_I-ks_{I}\phi$
is a constant independent of the index $I$.
In other words,
the product $q_I\wt q_I$ takes two values
according to $s_I$.
We can define ``meson operators'' $M$ and $M'$ by
\begin{equation}
M=q_i\wt q_i,\quad
M'=q_{i'}\wt q_{i'}.
\label{mmp}
\end{equation}
Now, we have $2r$ complex variables $q_I$ and $\wt q_I$ constrained by
(\ref{mmp}).
$\phi_a$ are dependent fields.
The number of independent complex variables is
$r+2$.
In addition to these, we need to take account of
the dual photon field $a$.
It is defined by solving the 
Gauss law constraint (\ref{gausslaw}) as
\begin{equation}
da=\sum_{a=1}^r k_aA_a.
\label{dualphoton}
\end{equation}
The dual photon field is combined with
the scalar field $\sigma$ in the diagonal $U(1)$ vector multiplet
to a complex scalar field belonging to a chiral multiplet.
It is convenient to define $e^{ia+\sigma}$.

Now we have $r+3$ independent complex variables.
We have to divide this space by complexified gauge symmetry $U(1)_C^{r-1}$
to obtain a complex $4$-dimensional moduli space.
Let us consider a gauge symmetry with parameter $\lambda_a$,
which transform the gauge fields by
\begin{equation}
\delta A_a=d\lambda_a.
\end{equation}
This transform the complex scalar fields
as
\begin{equation}
q_I\rightarrow e^{i\beta_I}q_I,\quad
e^{ia+\sigma}\rightarrow e^{-ik\sum_I s_I\beta_I}e^{ia+\sigma},
\end{equation}
where we defined
\begin{equation}
\beta_I=\lambda_{L(I)}-\lambda_{R(I)}.
\label{betai}
\end{equation}
By definition, parameters $\beta_I$ are constrained by
\begin{equation}
\sum_{I=1}^r\beta_I=0.
\end{equation}
Let us rewrite the parameters $\beta_I$ by $\varphi$, $\theta_i$, $\theta_{i'}$ as
\begin{equation}
\beta_i = {\varphi \over p} + \theta_i, \quad 
\beta_{i'} = - {\varphi \over q} + \theta_{i'},
\label{bpt}
\end{equation}
where $\theta_i$ and $\theta_{i'}$ satisfy
\begin{equation}
\sum_{i=1}^p\theta_i=\sum_{i'=1}^q\theta_{i'}=0.
\end{equation}
Then, the gauge transformation becomes
\begin{equation}
q_i\rightarrow e^{i\varphi/p}e^{i\theta_i}q_i,\quad
q_{i'}\rightarrow e^{-i\varphi/q}e^{i\theta_{i'}}q_{i'},\quad
e^{ia+\sigma}\rightarrow e^{-ik\varphi}e^{ia+\sigma}
\end{equation}
We can fix the continuous part of this gauge symmetry
by
\begin{equation}
e^{ia+\sigma}=1,
\label{phifix}
\end{equation}
and
\begin{equation}
q_{i=1}=\cdots =q_{i=p},\quad
q_{i'=1}=\cdots=q_{i'=q}.
\label{hip}
\end{equation}
(\ref{phifix}) fixes $\varphi$ transformation and
two equations in (\ref{hip}) fix the $\theta_i$ and $\theta_{i'}$
transformations.
If (\ref{hip}) hold,
the relations in (\ref{mmp})
guarantee
\begin{equation}
\wt q_{i=1}=\cdots =\wt q_{i=p},\quad
\wt q_{i'=1}=\cdots=\wt q_{i'=q}.
\label{hipt}
\end{equation}
After the gauge fixing, we have four independent complex variables.
We introduce the coordinates $z_m$ ($m=1,2,3,4$)
in the Higgs branch moduli space by
\be
z_1=q_i,\quad
z_2=\wt q_i^*,\quad
z_3=q_{i'},\quad
z_4=\wt q_i^*.
\label{zidef}
\ee
Even after the gauge fixing above,
we still have residual gauge symmetry with
the parameters
\be
\beta_i= {2\pi N \over kp} + {2\pi m \over p}, \quad 
\beta_{i'}= - {2\pi N \over kq} + {2\pi n \over q},
\ee
where $N$, $m$, $n$ are arbitrary integers. 
Due to this residual gauge symmetry
the global rotations
\begin{equation}
\exp\frac{2\pi i}{p}P,\quad
\exp\frac{2\pi i}{q}P',\quad
\exp (2\pi i P_M).
\label{threegen}
\end{equation}
are gauge equivalent to $1$,
$P$ and $P'$ are the generators of $U(1)$ and $U(1)'$, respectively,
and their action on the coordinates are shown in Table \ref{coords}.
$P_M$ is the linear combination of $P$ and $P'$;
\begin{equation}
P_M=\frac{1}{kq}P'-\frac{1}{kp}P.
\label{pmdef}
\end{equation}
The shift generated by $P_M$ is gauge equivalent
to the shift of dual photon field up to the gauge symmetry
associated with the parameter $\varphi$,
and we regard $P_M$ as the M-momentum.

By taking account of the discrete residual gauge symmetry (\ref{threegen}),
we obtain the moduli space
\begin{equation}
M_{p,q,k}=
((\CC^2/\ZZ_p)\times(\CC^2/\ZZ_q))/\ZZ_k.
\label{modsp}
\end{equation}

We summarize the action of global symmetries to the coordinates $z_m$
in Table \ref{coords}.
$T_3$ and $T_3'$ are the Cartan generators of $SU(2)$ and $SU(2)'$, respectively.
For convenience, we also include the weights $h_m$ ($m=1,\ldots,4$)
used in \S\ref{intro.sec} in the table.
\begin{table}[htb]
\caption{Actions of generators of global symmetries on the coordinates $z_1$, $z_2$, $z_3$, and $z_4$ are shown.}
\label{coords}
\begin{center}
\begin{tabular}{c|cccc|cccc}
\hline
\hline
& $T_3$ & $T_3'$ & $P$ & $P'$ & $h_1$ & $h_2$ & $h_3$ & $h_4$ \\
\hline
$z_1$ & $+1/2$ & $0$ & $+1$ & $0$ & $+1/2$ & $+1/2$ & $-1/2$ & $+1/2$ \\
$z_2$ & $-1/2$ & $0$ & $+1$ & $0$ & $-1/2$ & $+1/2$ & $+1/2$ & $+1/2$ \\
$z_3$ & $0$ & $+1/2$ & $0$ & $+1$ & $-1/2$ & $+1/2$ & $-1/2$ & $-1/2$ \\
$z_4$ & $0$ & $-1/2$ & $0$ & $+1$ & $+1/2$ & $+1/2$ & $+1/2$ & $-1/2$ \\
\hline
\end{tabular}
\end{center}
\end{table}
The charges $h_m$ are related to $P$, $P'$, $T_3$, and $T_3'$
by
\begin{equation}
h_1=T_3-T_3',\quad
h_2=\frac{1}{2}(P+P'),\quad
h_3=-(T_3+T_3'),\quad
h_4=\frac{1}{2}(P-P').
\end{equation}

\subsection{Wilson lines and fivebrane linking numbers}\label{linking.sec}
By restricting the orbifold $M_{p,q,k}$
on the sphere of the unit radius, we obtain
the internal space $X_7$ of the dual geometry
\begin{equation}
X_7 = \left({\bf S}^7/(\ZZ_p \times \ZZ_q)\right)/\ZZ_k. 
\label{x7}
\end{equation}
$\ZZ_k$ freely acts on the sphere and does not generate
fixed points, while
$\ZZ_p$ and $\ZZ_q$ generate three-dimensional loci of A-type singularities.
We denote the singular loci associated with
the $\ZZ_p$ and $\ZZ_q$ orbifoldings by
$\cSU$ and $\cST$, respectively.
$\cSU$ is $A_{p-1}$ singularity and an $SU(p)$ vector multiplet lives on it.
Similarly, on the other locus $\cST$, $SU(q)$ vector multiplet lives.
We define the gauge groups
\begin{equation}
G_\cS=G_{\cSU}\times G_{\cST}=SU(p)\times SU(q)
\end{equation}
and their Cartan parts
\begin{equation}
H_\cS=H_{\cSU}\times H_{\cST}=U(1)^{p-1}\times U(1)^{q-1}
\end{equation}
for later convenience.

It is often convenient to represent $X_7$ as a
$T^2$-fibration over a certain $5$-manifold by using the global
symmetry $U(1)\times U(1)'$ to define fibers as its orbits.
Then the loci $\cSU$ and $\cST$ are subsets of the base space
on which one cycle of the toric fiber shrinks.
(See \cite{Imamura:2008ji} for detailed description of the
structure of $X_7$.)
When we blow up the singularities, $\cSU$ and $\cST$ split into $p$ loci
$\cS_{Ui}$ and $q$ loci $\cS_{Ti'}$, respectively.
We here use indices $i$ and $i'$ just as for hypermultiplets.
The reason for this becomes clear shortly.
Each of the loci can be regarded as a brane which supports $U(1)$ vector multiplet
on its worldvolume.
(Precisely speaking, these $U(1)$ are not independent because the
gauge groups on the loci are not $U$ but $SU$.)
Topology of the loci $\cSU$ and $\cST$ are $S^3/\ZZ_{kq}$ and $S^3/\ZZ_{kp}$,
respectively.
Both the orbifold groups are generated by the third generator in
(\ref{threegen}).
Associated with the fundamental groups
$\pi_1(\cSU)=\ZZ_{kq}$ and $\pi_1(\cST)=\ZZ_{kp}$,
we have in general non-trivial Wilson lines
\begin{equation}
\diag(e^{2\pi i\eta_1},\ldots,e^{2\pi i\eta_p})\in U(p),\quad
\diag(e^{2\pi i\eta_1'},\ldots,e^{2\pi i\eta_q'})\in U(q),
\label{wlines}
\end{equation}
where each diagonal component of these Wilson lines corresponds to each of singular
loci $\cS_{Ui}$ or $\cS_{Ti'}$.
Note that these are not elements of $SU(p)$ and $SU(q)$ because
we do not impose the condition that their determinants are one.
This does not cause any problem because there are no particles
coupling to the $U(1)$ part.
$\eta_i$ and $\eta_{i'}$ must be quantized by
\begin{equation}
\eta_i\in\frac{1}{kq}\ZZ,\quad
\eta_{i'}\in\frac{1}{kp}\ZZ.
\label{etaquant}
\end{equation}
When we later compute the contribution of
twisted sectors to a multi-particle index,
we should take account of the momentum shift due to these Wilson lines.

To compare the multi-particle index with the gauge theory index,
we need to relate the Wilson lines to data of the gauge theory.
For this purpose, it is convenient to interpolate the M2-brane background
$M_{p,q,k}$ and the Chern-Simons theory by a type IIB brane system on which
the Chern-Simons theory is realized.
Let us consider $N$ coincident D3-branes wrapped around ${\bf S}^1$.
If the size of the ${\bf S}^1$ is small, 
the theory realized on the D3-brane worldvolume becomes effectively
three-dimensional.
We can realize $U(N)^r$ gauge group by introducing
$r$ fivebranes intersecting with the D3-brane worldvolume at
distinct points.
In type IIB theory fivebranes are characterized by two charges:
the NS5 charge and the D5 charge.
To realize ${\cal N}=4$ Chern-Simons theory with $p$ untwisted and $q$ twisted
hypermultiplets, we use $p$ NS5-branes
and $q$ $(k,1)$-fivebranes, and place them around
the ${\bf S}^1$ according to the quiver diagram.
$p+q=r$ hypermultiplets arise from massless modes of open
strings stretched between two adjacent intervals
on the both sides of the corresponding fivebrane.
The Chern-Simons terms (\ref{scs}) with levels (\ref{kkss})
are induced by the
boundary coupling at the ends of the intervals of
D3-branes\cite{Kitao:1998mf,Bergman:1999na}.

By the T-duality transformation along ${\bf S}^1$
and M-theory lift, this brane system is transformed into
$N$ M2-branes in the M-theory background $M_{p,q,k}$, and
its gravity dual is AdS$_4\times X_7$.
Through this duality chain, NS5-brane $i$ and $(1,k)$-fivebrane $i'$
are mapped to singular loci $\cS_{Ui}$ and $\cS_{Ti'}$, respectively.
We have already assumed implicitly this correspondence between
the fivebranes and singular loci when we used indices $i$ and $i'$ to
label the singular loci.

In \cite{Witten:2009xu}, the relation between the set of Wilson lines
and the structure of the brane system
is studied in detail for the case of $k=1$,
and it is shown that the Wilson lines are determined
by the fivebrane linking numbers.
The relation is easily generalized for general $k$.
To define the linking numbers,
we first need to choose one vertex in the circular quiver diagram
to cut the diagram at the vertex to make it linear.
We represent the reference vertex by $a=\bullet$.
For NS5-brane $i$ and $(1,k)$-fivebrane $i'$,
the linking numbers are defined by
\begin{equation}
l_i=\delta N_i+k\sum_{i<j'<\bullet}1,\quad
l_{i'}=\delta N_{i'}-k\sum_{\bullet<j<i'}1,
\label{linking}
\end{equation}
where $\delta N_I=N_{L(I)}-N_{R(I)}$ represents
the number of D3-branes ending on the fivebrane $I$.
When $k=1$ these reduces to those given in \cite{Witten:2009xu}.
We inserted $k$ so that these linking numbers are invariant in
the brane creation processes\cite{Hanany:1996ie}.
By definition, these numbers are integers depending on the reference point.
With these linking numbers, the Wilson line parameters are given by
\begin{equation}
\eta_i=\frac{1}{kq}l_i,\quad
\eta'_{i'}=\frac{1}{kp}l'_{i'}.
\end{equation}
These relations for $k=1$ are given in \cite{Witten:2009xu}.
We generalize them by inserting factor $k^{-1}$ so that
these relations are consistent with the quantization (\ref{etaquant}).

In this work, we only consider the case of $\delta N_I=0$,
and the linking numbers are multiples of $k$.

\section{Gravity side}\label{grav.sec}
The indices we consider in this paper are defined by
\begin{equation}
I(x,z,z')
=\tr\left[(-)^F
e^{-\beta'\{Q,S\}}x^{2(\Delta+j_3)}z^Pz'^{P'}\right].
\label{ixzz}
\end{equation}
where $Q$ is a certain component of the supercharge and $S$ is its
Hermitian conjugate.
On the gauge theory side, the trace is taken over all gauge invariant operators.
This index does not depend on $\beta'$, and only operators
saturating the BPS bound
\begin{equation}
\{Q,S\}=\Delta-j_3-(T_3+T_3')\geq0
\end{equation}
contribute.
We choose $Q$ so that
$h_3=-(T_3+T_3')$ is the R-charge
rotating $Q$.
The global symmetries commuting with this R-charge is
generated by $P$, $P'$, and $T_3-T_3'$.
Among these three $U(1)$ symmetries,
the last one
is broken when we deform the theory by adding $Q$-exact kinetic terms
for the vector multiplets for the purpose of taking the weak coupling limit.
This is the reason why we insert chemical potentials only for the
charges $P$ and $P'$.
We compare this index with the corresponding multi-particle index
for M-theory in the dual geometry AdS$_4\times X_7$.

On the gravity side, the single-particle index is given as the sum of two different origins.
One is the contribution of bulk particles,
and the other is that of the twisted sectors,
which are localized at the fixed loci in the orbifold.

\subsection{Bulk sector}
In this subsection, we discuss the bulk sector.
In general,
the index for bulk particles in an
orbifold ${\bf S}^7/\Gamma$ can be obtained
from the index for ${\bf S}^7$ by
the projection which leaves modes invariant under the orbifold action.
The single-particle index for bulk gravitons in AdS$_4\times{\bf S}^7$
is given in (\ref{igrav0}).

In the case of $X_7$ given in (\ref{x7}),
$\Gamma$ is generated by
the three generators in
(\ref{threegen}).
To obtain $\Gamma$-invariant part of the index,
let us first rewrite the index (\ref{igrav0})
as a function of $x$, $z$, and $z'$.
Because $z^Pz'^{P'}=(zz')^{h_2}(z/z')^{h_4}$, we can change the variables
by substituting
\begin{equation}
y_1=1,\quad
y_2=zz',\quad
y_3=\frac{z}{z'},
\label{yandz}
\end{equation}
into (\ref{igrav0}).
We obtain
\begin{equation}
I^{\rm grav}(x,z,z')=
\frac{
(\frac{1}{z}+z'+z+\frac{1}{z'})(x-x^7)
+(2+\frac{z'}{z}+\frac{1}{zz'}+zz'+\frac{z}{z'})(x^6-x^2)
}{
(1-x^4)
(1-xz')
(1-xz)
(1-x/z')
(1-x/z)}.
\label{igrav}
\end{equation}
We expand this index with respect to
$z$ and $z'$ as
\begin{equation}
I^{\rm grav}(x,z,z')
=\sum_{P,P'}I_{P,P'}^{\rm grav}(x)z^Pz'^{P'}.
\end{equation}
The coefficients $I_{P,P'}^{\rm grav}(x)$ are given by
\begin{equation}
I_{P,P'}^{\rm grav}(x)
=
(1-\delta_{P,0}\delta_{P',0})x^{|P|+|P'|}
+\delta_{P,0}\frac{x^{|P'|+2}}{1-x^4}
+\delta_{P',0}\frac{x^{|P|+2}}{1-x^4}.
\label{gravmn}
\end{equation}
The charges $P$ and $P'$ in $\Gamma$ invariant terms
must satisfy
\begin{equation}
\frac{1}{p}P\in\ZZ,\quad
\frac{1}{q}P'\in\ZZ,\quad
P_M\in\ZZ.
\end{equation}
The general solution to these conditions
is
\begin{equation}
P=pa,\quad
P'=q(a+kb),\quad
a,b\in\ZZ.
\label{kquantiz}
\end{equation}
The integer $b$ is equal to the M-momentum $P_M$.
The single-particle index for the bulk gravitons
in ${\bf S}^7/\Gamma$
is given by
\begin{equation}
I^{\cal B}(x,z,z')
=\sum_{a,b=-\infty}^\infty I_{pa,q(a+kb)}^{\rm grav}(x)z^{pa}z'^{q(a+kb)}.
\end{equation}

\subsection{Twisted sectors}\label{twisted.sec}
As we have already mentioned,
the internal space $X_7$ includes
two fixed loci $\cSU$ and $\cST$,
and we should take account of the twisted sectors associated with these.
The two sectors can be treated in parallel ways, and
we first consider the contribution of the $\cSU$ sector in detail.
Because $\cSU$ is the $A_{p-1}$ type singularity,
we expect that there exists an $SU(p)$ vector multiplet localized on
the locus.
With the coordinates defined in (\ref{zidef}), $\cSU$ is given by
$z_1=z_2=0$,
and is spanned by two complex coordinates $z_3$ and $z_4$ constrained
by
\begin{equation}
|z_3|^2+|z_4|^2=1.
\end{equation}
This equation together with the identification by the $\ZZ_{kq}$ generated by
the third generator in (\ref{threegen}) defines the Lens space ${\bf S}^3/\ZZ_{kq}$.
Because this orbifold does not have fixed points,
we can obtain the single-particle index for a vector multiplet
in this manifold by the $\ZZ_{kq}$ projection from the
index for the covering space ${\bf S}^3$.

The component fields in a vector multiplet in $\cSU$
do not carry the charge $P$.
Thus, the index should be the function only of $x$ and $z'$,
and is independent of $z$.
We propose the single-particle index
\begin{eqnarray}
I^{\rm vec}(x,z')
&=&\frac{x^2}{1-x^4}\left(1+\frac{xz'}{1-xz'}+\frac{x/z'}{1-x/z'}\right)
\nonumber\\
&=&\sum_{P'=-\infty}^\infty I_{P'}^{\rm vec}(x)z'^{P'},
\label{vecons3}
\end{eqnarray}
for a single $U(1)$ vector multiplet in ${\bf S}^3$,
where the coefficients in the $z'$ expansion are given by
\begin{equation}
I_m^{\rm vec}(x)=\frac{x^{|m|+2}}{1-x^4}.
\label{vecm}
\end{equation}
This single-particle index should be directly derived
from the analysis of Kaluza-Klein spectrum
of a vector multiplet on ${\bf S}^3$.
We leave such analysis for future work,
and use this as a starting point of the analysis of the
twisted sectors.
Once we accept that the single-particle index for ${\bf S}^3$ is given by (\ref{vecons3}),
the index for orbifold ${\bf S}^3/\ZZ_{kq}$ is obtained by
the projection which leaves only $\ZZ_{kq}$ invariant modes.

When we consider the single-particle index of the covering space of
the other locus $\cST$, we should replace the variable $z'$
in (\ref{vecons3}) by $z$.
Namely, it is $I^{\rm vec}(x,z)$.

The procedure of the $\ZZ_{kq}$ and $\ZZ_{kp}$ projections is
similar to what we have done for the bulk sector.
An important difference is that
in general there exist
non-trivial Wilson lines coupling to the vector multiplets
in the twisted sectors.
Before considering the projection for single-particle states,
let us consider that for a general multi-particle state.
Let $\rho_i$ and $\rho_{i'}$ be the $H_{\cSU}$ and $H_{\cST}$ charges
of the multi-particle state.
They are the sum of charges of constituent particles in the state.
Because every particle belongs to the adjoint representation of $G_\cS$,
these charges satisfy
\begin{equation}
\sum_{i=1}^p\rho_i=\sum_{i'=1}^q\rho_{i'}=0.
\label{sumrho}
\end{equation}
When we act an element of the orbifold group
which rotates the cycles in $\cSU$ and $\cST$ by $r$ and $s$ times, respectively,
the state picks up the phase
\begin{equation}
2\pi i\left(
r\sum_{i=1}^p\rho_i\eta_i
+s\sum_{i'=1}^q\rho_{i'}\eta_{i'}\right),
\end{equation}
and this must be canceled by the phase factor associated with the momentum.
Because
$(r,s)=(0,-k)$, $(k,0)$, and $(1,1)$
for the three generators in (\ref{threegen}),
the cancellation of the phases requires
\begin{equation}
\exp\left(\frac{2\pi i}{p}P\right)
=\exp\left(2\pi ik\sum_{i'=1}^q\rho_{i'}\eta_{i'}\right),
\end{equation}
\begin{equation}
\exp\left(\frac{2\pi i}{q}P'\right)
=\exp\left(-2\pi ik\sum_{i=1}^p\rho_i\eta_i\right),
\end{equation}
\begin{equation}
\exp(2\pi i P_M)
=\exp\left(-2\pi i\sum_{i=1}^p\rho_i\eta_i
-2\pi i\sum_{i'=1}^q\rho_{i'}\eta_{i'}\right).
\end{equation}
$P$ and $P'$ satisfying these conditions are given by
\begin{equation}
P=p\left(a+k\sum_{i'=1}^q\rho'_{\alpha i'}\eta'_{i'}\right),\quad
P'=q\left(a+kb-k\sum_{i=1}^p\rho_{\alpha i}\eta_i\right),\quad
a,b\in\ZZ,
\label{ppselect}
\end{equation}
and then the M-momentum $P_M$ is
\begin{equation}
P_M
=b-\sum_{i=1}^p\rho_{\alpha i}\eta_i-\sum_{i'=1}^q\rho'_{\alpha i'}\eta'_{i'},
\quad
b\in\ZZ.
\label{pmselect}
\end{equation}
Unlike the case of bulk sector, the M-momentum $P_M$ is not always an integer.
These conditions are imposed on any multi-particle states,
including single-particle states.
Actually, we obtain the momenta (\ref{kquantiz}) for
bulk single-particle states
by simply setting $\rho_i=\rho_{i'}=0$ in
(\ref{ppselect}).

For a single-particle state in the twisted sector on the locus $\cSU$,
$\rho_{i'}=P=0$.
This implies that $a$ in
the first equation in (\ref{ppselect}) vanishes,
and the second equation gives the momentum
\begin{equation}
P'=kq\left(b-\sum_{i=1}^p\rho_{\alpha i}\eta_i\right),\quad
b\in\ZZ.
\label{pprime}
\end{equation}
Let $\{\rho_{\alpha i}\}=\vec\rho_\alpha$ be the charge vector
for an $SU(p)$ vector multiplet living in the locus $\cSU$.
$\alpha=1,\ldots,p^2-1$ is the adjoint index of $SU(p)$.
These vectors are
nothing but the weight vectors for the adjoint representation
of $SU(p)$.
The single particle index for vector multiplets
in the locus $\cSU$ is
\begin{eqnarray}
I^{\cSU}(x,z';\vec t)
&=&\sum_{\alpha=1}^{p^2-1}\sum_{b=-\infty}^\infty
I^{\rm vec}_{kq(b-\vec\rho_\alpha\cdot\vec\eta)}(x)z'^{kq(b-\vec\rho_\alpha\cdot\vec\eta)}
t_1^{\rho_{\alpha 1}}\cdots
t_p^{\rho_{\alpha p}}
\nonumber\\
&=&\sum_{\vec\rho}\deg(\vec\rho)\sum_{b=-\infty}^\infty
I^{\rm vec}_{kq(b-\vec\rho\cdot\vec\eta)}(x)z'^{kq(b-\vec\rho\cdot\vec\eta)}
t_1^{\rho_1}\cdots
t_p^{\rho_p},
\end{eqnarray}
where $\deg(\vec\rho)$ is the degeneracy for the
adjoint representation at $\vec\rho$ in the $SU(p)$ root lattice.
Namely,
\begin{equation}
\deg(\vec\rho)=\left\{
\begin{array}{cl}
1 & (|\vec\rho|^2=2)\\
p-1 & (\vec\rho=0) \\
0 & (\mbox{others})
\end{array}\right.
\label{degen}
\end{equation}
We here introduced new chemical potentials $\vec t=(t_1,\ldots, t_p)$
for the $H_{\cSU}$ charges $\vec\rho$.

The single-particle index for the $SU(q)$ vector multiplet
localized in the locus
$\cST={\bf S}^3/\ZZ_{kp}$ is obtained in the same way.
Because $\rho_i=P'=0$,
the projection restrict the value of the momentum $P$ as
\begin{equation}
P=kp\left(-b+\sum_{i'=1}^q\rho'_{\alpha i'}\eta'_{i'}\right),\quad
b\in\ZZ.
\label{zpquantize}
\end{equation}
The single-particle index for the $SU(q)$ vector multiplet
in $\cST$ is given by
\begin{eqnarray}
I^{\cST}(x,z;\vec t')
&=&\sum_{\alpha=1}^{q^2-1}\sum_{a=-\infty}^\infty
I^{\rm vec}_{kp(a+\vec\rho'_\alpha\cdot\vec\eta)}(x)z^{kq(p+\vec\rho'_\alpha\cdot\eta_\alpha')}
t_1'^{\rho_{\alpha 1}'}
\cdots t_q'^{\rho_{\alpha q}'}
\nonumber\\
&=&\sum_{\vec\rho'}\deg(\vec\rho')\sum_{a=-\infty}^\infty
I^{\rm vec}_{kp(a+\vec\rho'\cdot\vec\eta)}(x)z^{kq(p+\vec\rho'\cdot\eta_\alpha')}
t_1'^{\rho_1'}
\cdots t_q'^{\rho_q'},
\end{eqnarray}
where we defined the degeneracy for the $SU(q)$ adjoint representation
similarly to (\ref{degen}).

Due to the constraint (\ref{sumrho})
these indices are invariant under the overall rescaling of $\vec t$
and $\vec t'$;
\begin{equation}
I^{\cSU}(x,z',c\vec t)
=I^{\cSU}(x,z',\vec t),\quad
I^{\cST}(x,z,c\vec t')
=I^{\cST}(x,z,\vec t').
\end{equation}

By summing up the contribution of the bulk and the twisted sectors,
we obtain
\begin{equation}
I^{\rm sp}(x,z,z';\vec t,\vec t')
=I^{\cal B}(x,z,z')
+I^{\cSU}(x,z';\vec t)
+I^{\cST}(x,z;\vec t').
\label{ispdef}
\end{equation}

In the following sections, we confirm that
the corresponding multi-particle index
\begin{equation}
I^{\rm mp}(x,z,z';\vec t,\vec t')
=\exp\sum_{n=1}^\infty\frac{1}{n}I^{\rm sp}(x^n,z^n,z'^n;\vec t^n,\vec t'^n)
\label{Imulti}
\end{equation}
is reproduced as the monopole index
on the field theory side.
What we will actually do in the following section is not to
derive the indices as functions of $\vec t$ and $\vec t'$
but to compute indices for various sectors specified by charges $(P_M,\vec\rho,\vec \rho')$
separately.
The index for each sector specified
by $(P_M,\vec\rho,\vec\rho')$
is extracted from (\ref{Imulti}) by
\begin{equation}
I^{\rm mp}(x,c^{-\frac{1}{kp}}z,c^{\frac{1}{kq}}z';\vec t,\vec t')=
\sum_{(P_M,\vec\rho,\vec\rho')} I^{\rm mp}_{(P_M,\vec\rho,\vec\rho')}(x,z,z')
c^{P_M}\vec t^{\vec\rho}
\vec t'^{\vec\rho'}.
\end{equation}
To pick up the part of specific M-momentum
we inserted an auxiliary variable $c$.
The summation with respect to the vectors $\vec\rho$ and $\vec\rho'$
are taken over the $SU(p)$ and $SU(q)$ root lattices.
Note that $P_M$ is not always integer.
The values $P_M$ can take are determined by
the equation (\ref{pmselect}), and depend on $\vec\rho$ and $\vec\rho'$.

The left and right hand sides in (\ref{ispdef})
are also expanded in a similar way, and we obtain
\begin{eqnarray}
I^{\rm sp}_{(P_M,\vec\rho,\vec\rho')}(x,z,z';\vec t,\vec t')
&=&\delta_{\vec\rho,\vec0}\delta_{\vec\rho',\vec0}
I_{P_M}^{\cal B}(x,z,z')
\nonumber\\
&&+\delta_{\vec\rho',\vec0}I^{\cSU}_{(P_M,\vec\rho)}(x,z')
\vec t^{\vec\rho}
\nonumber\\
&&+\delta_{\vec\rho,\vec0}I^{\cST}_{(P_M,\vec\rho')}(x,z)
\vec t'^{\vec\rho'},
\label{ispppp}
\end{eqnarray}
where indices on the right hand side are defined by
\begin{eqnarray}
I_{P_M}^{\cal B}(x,z,z')
&=&\sum_{a=-\infty}^\infty I^{\rm grav}_{pa,q(a+kP_M)}(x)z^{pa}z'^{q(a+kP_M)},\\
I^{\cSU}_{(P_M,\vec\rho)}(x,z',\vec t)
&=&\deg(\vec\rho)I^{\rm vec}_{kqP_M}(x)z'^{kqP_M}\vec t^{\vec\rho},\\
I^{\cST}_{(P_M,\vec\rho')}(x,z,\vec t')
&=&\deg(\vec\rho')I^{\rm vec}_{-kpP_M}(x)z^{-kpP_M}\vec t'^{\vec\rho'}.
\end{eqnarray}

\section{Gauge theory side}\label{gauge.sec}
\subsection{Gauge theory index}
The gauge theory index $I^{\rm gauge}$ which we study is defined by
\begin{equation}
I^{\rm gauge}(x,z,z')=\tr\left[
(-)^Fe^{-\beta'\{Q,S\}}x^{2(\Delta+j_3)}z^Pz^{P'}
\right].
\label{igauge}
\end{equation}
This is evaluated by the radial quantization method\cite{Borokhov:2002ib,Borokhov:2002cg}.
The procedure to compute
this index is essentially the same as the case of the
ABJM model,
which is explained in \cite{Kim:2009wb} in detail.

By a conformal transformation, a local operator in $\RR^3$ is
mapped to a state in the Fock space of the conformal field theory
defined in ${\bf S}^2\times \RR$.
The trace over all operators is replaced by the path integral
in the compact three-dimensional space ${\bf S}^2\times{\bf S}^1$,
where ${\bf S}^1$ is the compactified time direction.
To carry out the path integral,
we need to take a weak coupling limit in such a way
that it does not change the index.
If the theory had continuous coupling constants,
we could take such a limit by sending them to zero.
In the theory we discuss, however, we do not have such
continuous parameters.
In the large $N$ limit with fixed 't~Hooft coupling
$\lambda=N/k$, $\lambda$ becomes effectively continuous,
and we can take the weak coupling limit $\lambda\rightarrow0$.
This procedure is used
in \cite{Choi:2008za} to compute
the neutral part of the index (\ref{n4sfindex}).
(We mean by the neutral part the contribution of
operators without magnetic charges.)
However, we cannot use the same procedure
because the monopole contributions are suppressed
in the large $k$ limit.

In this paper, we keep the level $k$ finite,
and take a weak coupling limit by adding
$Q$-exact terms to the action.
We can realize kinetic terms of vector and hypermultiplets
as $Q$-exact terms, and
adding such terms to the action does not affect the index
because only states eliminated by $Q$ contribute to the index.
We can take the weak coupling limit
by sending the coefficients of the $Q$-exact terms to infinity.
In such a limit,
we can treat all fields as free fields, and
the saddle point approximation gives the exact result.

Because we want to take account of monopole operators,
we should consider all backgrounds
with magnetic flux through the ${\bf S}^2$.
We assume that only 
Goddard-Nuyts-Olive (GNO) monopoles\cite{Goddard:1976qe} contribute
to the index
as saddle points in the
path integral.
GNO monopoles are superposition of Dirac monopoles for the
Cartan part of the gauge group.
For every vertex $a$ in the quiver diagram we have $N$ $U(1)$ subgroups.
We label them by color indices $s$, $t$, $\ldots$.
Let $m_{as}\in\ZZ$ be the magnetic charge
of the GNO monopole for the $s$-th $U(1)$ subgroup
of $U(N)_a$.
We should consider all possible charges parameterized by $rN$ integers
$\{m_{as}\}$, and the total index is given as the summation over all
monopole charges
\begin{equation}
I^{\rm gauge}(x,z,z';\vec\tau)=\sum_{\{m_{as}\}}I_{\{m_{as}\}}(x,z,z')\prod_{a=1}^n\tau_a^{m_a},
\label{gaugeindex}
\end{equation}
where we introduced chemical potentials $\tau_a$ for
the magnetic charges $m_a$ defined by
\begin{equation}
m_a=\sum_{s=1}^Nm_{as}.
\end{equation}
These are the gauge invariant monopole charges introduced in (\ref{madef}).

In order to compare the gauge theory index 
(\ref{gaugeindex}) with the multi-particle index
(\ref{Imulti})
derived on the
gravity side,
we need to find the relation between magnetic charges $m_a$ and
the variables $(P_M,\vec\rho,\vec\rho')$.
We discuss this relation in the next subsection.
Here we focus on the way to compute the index for
each sector specified by the magnetic charges.

In the weak coupling limit, we can expand fields in the theory by harmonic functions in ${\bf S}^2$ with magnetic flux
(monopole harmonics), and the path integral reduces
to integrals for
the infinite number of modes.
The Gaussian integrals associated with non-zero modes
can be easily performed,
and we are left with the expression with
integrals with respect to the holonomy
around the compact time direction
\begin{equation}
\diag(e^{i\alpha_{a1}},\ldots,
e^{i\alpha_{as}},\ldots
e^{i\alpha_{aN}})
\in U(N)_a
\end{equation}
in the Cartan part of the gauge group.
The expression after the integration over the massive modes is
\begin{eqnarray}
I_{\{m_{as}\}}(x,z,z')
&=&
x^{2\epsilon_0(\{m_{as}\})}
\left(\prod_{a=1}^{r}\prod_{s=1}^N \int\frac{d\alpha_{as}}{2\pi}\right)
\exp\left(i\sum_{a=1}^{r}\sum_{s=1}^N k_am_{as}\alpha_{as}\right)
\nonumber\\&&\times
\prod_{a=1}^r\prod_{s,t=1}^N\exp\left[
\sum_{n=1}^\infty\frac{1}{n}
f_{ast}^{\rm vec}(\{m_{as}\};x^n,e^{in\beta_{ast}})
\right]
\nonumber\\&&\times
\prod_{I=1}^r\prod_{s,t=1}^N\exp\left[
\sum_{n=1}^\infty\frac{1}{n}
f_{Ist}^{\rm hyp}(\{m_{as}\};x^n,(z_Ie^{i\beta_{Ist}})^n)
\right].
\label{totindex}
\end{eqnarray}
See \cite{Kim:2009wb,Kim:2009ia} for a detailed derivation in the case of ABJM model.
Generalization to N=4 theories is straightforward.
$\epsilon_0$ is the zero point energy
due to the vacuum polarization in ${\bf S}^2$,
and is given by
\begin{equation}
\epsilon_0(\{m_{as}\})=-\frac{1}{2}\sum_{a=1}^{r}\sum_{s,t=1}^N|m_{as}-m_{at}|
    +\frac{1}{2}\sum_{I=1}^{r}\sum_{s,t=1}^N|m_{L(I)s}-m_{R(I)t}|,
\label{e00}
\end{equation}
where the first and the second terms are the
contribution of vector and hyper multiplets, respectively.
$f_{ast}^{\rm vec}$ and $f_{Ist}^{\rm hyp}$
are contributions of oscillators in the vector multiplet $V_a$ and hyper
multiplet $H_I$, respectively, and given by
\begin{eqnarray}
f_{ast}^{\rm vec}(\{m_{as}\};x,e^{i\beta_{ast}})&=&
-(1-\delta_{st})x^{2|m_{as}-m_{at}|}e^{i\beta_{ast}},\\
f_{Ist}^{\rm vec}(\{m_{as}\};x,z_Ie^{i\beta_{Ist}})&=&
\frac{x^{2|m_{L(I)s}-m_{R(I)t}|+1}}{1+x^2}
\left(e^{i\beta_{Ist}}z_I
+\frac{1}{e^{i\beta_{Ist}}z_I}\right).
\end{eqnarray}
$z_I$ is defined by
\begin{equation}
z_I=\left\{
\begin{array}{ll}
z & \mbox{for $s_I=0$}\\
z' & \mbox{for $s_I=1$}
\end{array}\right.
\end{equation}
$\beta_{ast}$ and $\beta_{Ist}$ are angular variables defined by
\begin{equation}
\beta_{ast}=\alpha_{as}-\alpha_{at},\quad
\beta_{Ist}=\alpha_{L(I)s}-\alpha_{R(I)t}.
\end{equation}
These angular variables are holonomies for the
components of $V_a$ or $H_I$ specified by the color indices
$s$ and $t$.

To obtain the gauge theory index which can be compared with the
graviton index, we should take the large $N$ limit.
This limit is taken by adding vanishing entries to
the monopole charges $\{m_{as}\}$.
For each $a$, the monopole charge is described by $N$ integers
$m_{as}$ ($s=1,\ldots,N$).
Let $M_a$ be the number of non-vanishing components among them.
When we take the large $N$ limit, we keep $M_a$ at ${\cal O}(1)$.

For this limit to be well defined,
the zero-point energy should not diverge in the limit.
This is indeed easily confirmed by rewriting (\ref{e00}) as
\begin{eqnarray}
\epsilon_0(\{m_{as}\}_*)&=&
-\frac{1}{2}\sum_{a=1}^{r}\sum_{s\in M_a}\sum_{t\in M_a}|m_{as}-m_{at}|
+\frac{1}{2}\sum_{a=1}^{r}\sum_{s\in M_{L(I)}}\sum_{t\in M_{R(I)}}|m_{L(I)s}-m_{R(I)t}|
\nonumber\\&&
+\frac{1}{2}\sum_{a=1}^{r}
    (2M_a-M_{a+1}-M_{a-1})\sum_{s\in M_a}|m_{as}|,
\label{e0pm}
\end{eqnarray}
where
$\{m_{as}\}_*$ is the collection of non-vanishing components
in $\{m_{as}\}$, and
$\sum_{s\in M_a}$ represents the summation over $M_a$ non-vanishing
components in the magnetic charges.
This expression is manifestly independent of $N$, and well behaves
in the large $N$ limit.

The integration with respect to angular variables $\alpha_{as}$ associated with
vanishing magnetic charges $m_{as}$ can be carried out
by introducing the variables $\lambda_{an}$ by
\begin{equation}
\lambda_{an}=\frac{1}{N-M_a}\sum_{s={M_a+1}}^N e^{in\alpha_{as}},\quad
n=\pm1,\pm2,\ldots.
\end{equation}
The exponential factors in the second and the third lines
in (\ref{totindex}) can be rewritten as a Gaussian factor including
\begin{equation}
\exp\left(-\sum_{n=1}^\infty\frac{1}{n}\sum_{a,b=1}^{r}\lambda_{an}M_{ab}(x^n,z^n,z'^n)\lambda_{b-n}
+\cdots\right)
\end{equation}
where $\cdots$ include the first and the zeroth order terms of $\lambda_{an}$,
and the matrix $M$ is
\begin{equation}
M(x,z,z')=\left(
\begin{array}{cccccc}
\ddots &&&&& -\frac{xz^{-1}_{r}}{1+x^2} \\
& 1 & -\frac{xz_{I-1}}{1+x^2} \\
& -\frac{xz_{I-1}^{-1}}{1+x^2} & 1 & -\frac{xz_I}{1+x^2} \\
&& -\frac{xz_I^{-1}}{1+x^2} & 1 & -\frac{xz_{I+1}}{1+x^2} \\
&&& -\frac{xz_{I+1}^{-1}}{1+x^2} & 1 &  \\
-\frac{xz_{r}}{1+x^2} &&&&& \ddots \\
\end{array}
\right).
\label{mdef}
\end{equation}
After the Gaussian integral with respect to $\lambda_{an}$,
we are left with the following expression
including the finite number of integrals.
\begin{equation}
I_{\{m_{as}\}}(x,z,z')=I^{(0)}(x,z,z')I_{\{m_{as}\}_*}^{(*)}(x,z,z'),
\end{equation}
where $I^{(0)}$ is the determinant factor associated with the
Gaussian integral of $\lambda_{an}$:
\begin{equation}
I^{(0)}(x,z,z')=\prod_{n=1}^\infty\frac{1}{\det M(x^n,z^n,z'^n)},
\label{izero}
\end{equation}
and $I_{\{m_{as}\}_*}^{(*)}$ is given by
\begin{eqnarray}
I^{(*)}_{\{m_{as}\}_*}(x,z,z')
&=&
\frac{x^{2\epsilon_0(\{m_{as}\}_*)}}{(\mbox{symmetry})}
\left(
\prod_{a=1}^{r}
\prod_{s=1}^{M_i}
\int\frac{d\alpha_{as}}{2\pi}
\right)
\exp\left(i\sum_{a=1}^{r}\sum_{s=1}^{M_a}k_am_{as}\alpha_{as}\right)
\nonumber\\&&
\times\prod_a\prod_{s,t}
\left[\exp\sum_{n=1}^\infty
{\bf f}_{ast}^{\rm vec}(\{m_{as}\};x^n,e^{in\beta_{ast}})
\right]
\nonumber\\&&
\times\prod_I\prod_{s,t}
\left[\exp\sum_{n=1}^\infty
{\bf f}_{Ist}^{\rm hyp}(\{m_{as}\};x^n,(z_Ie^{i\beta_{Ist}})^n)
\right].
\label{ipom}
\end{eqnarray}
For later convenience we divide the magnetic charges $\{m_{as}\}_*$
into $\{m_{as}\}_+$, the collection of positive charges,
and $\{m_{as}\}_-$, the collection of negative charges,
and represent each of $\{m_{as}\}_+$ and $\{m_{as}\}_-$
as a set of $r$ Young diagrams.
We also introduce $M_a^+$ ($M_a^-$),
the number of positive (negative) components in $m_{as}$ for
each $a$.
The symmetry factor in
(\ref{ipom})
is the product of the symmetry factors of
the $2r$ Young diagrams.
The symmetry factor for a single Young diagram
is defined as the product of the factorial of the
number of the lines with the same length in the Young diagram.
For example, the symmetry factor for $\Y(3,1,1)$ and $\Y(4,4,1,1,1)$ are
$1!2!$ and $2!3!$, respectively.

${\bf f}^{\rm vec}_{ast}$ and ${\bf f}^{\rm hyp}_{Ist}$ are given by
\begin{eqnarray}
{\bf f}^{\rm vec}_{ast}(\{m_{as}\};x,e^{i\beta_{ast}})&=&
\left(
-(1-\delta_{st})x^{2|m_{as}-m_{at}|}
+x^{2(|m_{as}|+|m_{at}|)}
\right)e^{i\beta_{ast}},
\label{fap}\\
{\bf f}^{\rm hyp}_{Ist}(\{m_{as}\};x,z_Ie^{i\beta_{Ist}})&=&
(x^{2|m_{L(I)s}-m_{R(I)t}|}-x^{2(|m_{L(I)s}|+|m_{R(I)t}|)})
\nonumber\\&&\times
\frac{x}{1+x^2}
\left(z_Ie^{i\beta_{Ist}}+\frac{1}{z_Ie^{i\beta_{Ist}}}
\right)
\label{fip}
\end{eqnarray}

We can easily see that $I_{\{m_{as}\}_*}^{(*)}$ is further factorized into
$I_{\{m_{as}\}_+}^{(+)}$ depending only on $\{m_{as}\}_+$ and
$I_{\{m_{as}\}_-}^{(-)}$ depending only on $\{m_{as}\}_-$.
To show the factorization of the zero-point energy contribution
$x^{2\epsilon_0}$, we divide the range of all the summations
of color indices $s$ and $t$ in (\ref{e0pm})
into two parts as
$\sum_{s\in M_a}=\sum_{s\in M_a^+}+\sum_{s\in M_a^-}$.
The first term in (\ref{e0pm}) is decomposed as
\begin{eqnarray}
&&
-\sum_{a=1}^{r}\sum_{s\in M_a^+}\sum_{t\in M_a^+}|m_{as}-m_{at}|
-\sum_{a=1}^{r}\sum_{s\in M_a^-}\sum_{t\in M_a^-}|m_{as}-m_{at}|
\nonumber\\
&&
-\sum_{a=1}^{r}\left(2M_a^-\sum_{s\in M_a^+}|m_{as}|+2M_a^+\sum_{s\in M_a^-}|m_{as}|\right)
\end{eqnarray}
In the first line, the contributions of $\{m_{as}\}_+$ and $\{m_{as}\}_-$ decouple from each other.
The two terms in the second line depend on both
$\{m_{as}\}_+$ and $\{m_{as}\}_-$,
and for the factorization, these terms should be canceled by other terms.
Actually, these terms are precisely canceled by the
mixed terms arising from the
$\sum2M_a\sum|m_{as}|$ term in the second line in
(\ref{e0pm}).
In this way, all the mixed terms cancel,
and the zero-point energy is represented as
\begin{equation}
\epsilon_0(\{m_{as}\}_*)=\epsilon_0(\{m_{as}\}_+)+\epsilon_0(\{m_{as}\}_-).
\end{equation}
The factorization of the second and the third lines in 
(\ref{ipom}) is shown by using the fact that
the factor in the form
$x^{2|m-m'|}-x^{2(|m|-|m'|)}$
appearing in (\ref{fap}) and (\ref{fip})
vanish when $m$ and $m'$ have opposite signatures.

Now we have shown that the
gauge theory index factorizes into the three parts:
\begin{equation}
I_{\{m_{as}\}}(x,z,z')=I^{(0)}(x,z,z')I_{\{m_{as}\}_+}^{(+)}(x,z,z')I_{\{m_{as}\}_-}^{(-)}(x,z,z').
\end{equation}
Because the summations over $\{m_{as}\}_+$ and $\{m_{as}\}_-$ are
independent in the large $N$ limit,
the total index also factorizes into three parts
\begin{equation}
I^{{\rm guage}}(x,z,z';\vec\tau)=I^{(0)}(x,z,z')I^{(+)}(x,z,z';\vec\tau)I^{(-)}(x,z,z';\vec\tau),
\end{equation}
where
$I^{(\pm)}$ $(x,z,z';\vec\tau)$ is defined by
\begin{equation}
I^{(\pm)}(x,z,z';\vec\tau)
=\sum_{\{m_{as}\}_\pm}I^{(\pm)}_{\{m_{as}\}_\pm}(x,z,z')
\tau_1^{m_1}\cdots\tau_r^{m_r}.
\end{equation}
We also define the index for a specific gauge invariant monopole
charges $\{m_a\}$ as the sum of contributions of all
the monopole backgrounds with the same $\{m_a\}$.
For example, the index for $\{m_a\}=\{2,2\}$ is
the sum of four contributions:
\begin{equation}
I^{(+)}_{\{2,2\}}
=I^{(+)}_{\{\Y(2),\Y(2)\}}
+I^{(+)}_{\{\Y(2),\Y(1,1)\}}
+I^{(+)}_{\{\Y(1,1),\Y(2)\}}
+I^{(+)}_{\{\Y(1,1),\Y(1,1)\}},
\end{equation}
where we used the Dynkin diagrams to represents
the charges $\{m_{as}\}_+$.

\subsection{Selection rules}
The integration with respcet to the angular variable $\alpha_{ia}$
leaves only terms
whose $P$ and $P'$,
the numbers of $z$ and $z'$ in the terms,
satisfy certain selection rules, 
which correspond to conditions of gauge-invariance of operators.
For an operator which carries at most the diagonal U(1) magnetic charge, 
selection rules are expected to be
\be
{1\over p}P \in \ZZ, \quad {1\over q} P'\in\ZZ,
\label{pqmult}
\ee
which means that such an operator is invariant under the residual gauge 
transformations (\ref{threegen}). 
For another operator which carries different magnetic charges, 
selection rules are considered to shift from (\ref{pqmult})\cite{Imamura:2009ur}.
In this subsection, we derive selection rules 
and show that these selection rules precisely reproduce the spectrum (\ref{ppselect})
of the Kaluza-Klein momenta derived on the gravity side.

Let us start from (\ref{totindex}).
For every vertex ($U(N)$ gauge group) $a$,
we have $N$ angular variables $\alpha_{as}$ ($s=1,\ldots,N$).
Instead of these,
let us take $\alpha_{a1}$ and $\beta_{a1s}$ ($s=2,\ldots,N$)
as $N$ independent angular variables.
We replace all $\alpha_{as}$ ($s\geq2$) in (\ref{totindex}) by $\alpha_{a1}-\beta_{a1s}$.
By this replacement,
the exponential factor including the levels $k_a$ becomes
\begin{equation}
\exp\left(i\sum_{a=1}^r\sum_{s=1}^Nk_am_{as}\alpha_{as}\right)
=\exp\left(i\sum_{a=1}^rk_am_a\alpha_{a1}\right)
\times\exp\left(-i\sum_{a=1}^r\sum_{s=2}^Nk_am_{as}\beta_{a1s}\right).
\end{equation}
The variables $\beta_{Ist}$ in $f_{Ist}^{\rm hyp}(x,z_Ie^{i\beta_{Ist}})$ become
\begin{equation}
\beta_{Ist}=\beta_{I11}-\beta_{L(I)1s}+\beta_{R(I)1t}.
\end{equation}
As a result, the parameter $z_I$ is always accompanied
by $e^{i\beta_{I11}}$.
After integrating out $\beta_{a1s}$ ($s\geq2$),
we obtain
\begin{equation}
I_{\{m_{as}\}}(x,z,z')
=
\left(\prod_{a=1}^{r}\int d\alpha_{a1}\right)
\exp\left(i\sum_{a=1}^{r}
k_am_a\alpha_{a1}\right)
f(z_Ie^{i\beta_{I11}}),
\label{intai1}
\end{equation}
where $f(z_Ie^{i\beta_{I11}})$ is a certain function of
$r$ variables $z_Ie^{i\beta_{I11}}$ ($I=1,\ldots,r$).
This function also depends on $x$, but we do not take
care about it here.

We now have $r$ angular variables $\alpha_{a1}$ to be integrated.
Instead of these,
let us use $\alpha_{\bullet 1}$ and $\beta_{I11}$ as $r$ independent variables,
where $a=\bullet$ is the reference vertex
we used to define the fivebrane linking numbers
in \S\ref{linking.sec}.
By definition $\beta_{I11}$ satisfy
\begin{equation}
\sum_{I=1}^{r}\beta_{I11}=0.
\end{equation}
To treat all $\beta_{I11}$ as independent variables,
we insert the $\delta$ function
\begin{equation}
\delta(\sum_{I=1}^r\beta_{I11})
=\sum_{d=-\infty}^\infty\exp(-id\sum_{I=1}^r\beta_{I11})
\end{equation}
into (\ref{intai1}).
($\delta(\theta)$ in this equation is the $\delta$-function for an
angular variable.
Namely it has the periodic support $\theta=2\pi n$.)
We rewrite the exponential factor in (\ref{intai1}) by using
\begin{equation}
\sum_{a=1}^{r}k_am_a\alpha_{a1}
=k\sum_{I=1}^rc_I\beta_{I11}
-k\alpha_{\bullet 1}\sum_{I=1}^rs_I\mu_I,
\end{equation}
where we defined
\begin{equation}
c_I=\sum_{\bullet<J<I}(s_I-s_J)\mu_J-m_\bullet s_I,
\end{equation}
and the relative magnetic charges
\begin{equation}
\mu_I=m_{L(I)}-m_{R(I)}.
\end{equation}
By definition $\mu_I$ satisfy
\begin{equation}
\sum_{I=1}^r\mu_I=0.
\label{summu0}
\end{equation}
We obtain
\begin{eqnarray}
I_{\{m_{as}\}}(x,z,z')
&=&
\sum_{d=-\infty}^\infty
\int d\alpha_{\bullet 1}\left(\prod_{I=1}^{r}\int d\beta_{I11}\right)
F(z_Ie^{i\beta_{I11}})
\nonumber\\&&
\times\exp
\left(i\sum_{I=1}^r(kc_I-d)\beta_{I11}
-k\alpha_{\bullet 1}\sum_{I=1}^rs_I\mu_I
\right).
\end{eqnarray}
The integration of $\alpha_{\bullet 1}$ gives the constraint
\begin{equation}
\sum_{I=1}^rs_I\mu_I=0,
\label{smuzero}
\end{equation}
imposed on $\mu_I$.
This is equivalent to (\ref{kmzero}).
It is convenient to divide the relative magnetic charges
$\mu_I$ into two sets
$\mu_i$ and $\mu_{i'}$
corresponding to two kinds of hypermultiplets.
(\ref{summu0}) and (\ref{smuzero}) are equivalent to the two constraints
\begin{equation}
\sum_{i=1}^p\mu_i=0,\quad
\sum_{i'=1}^q\mu_{i'}=0,
\end{equation}
and they form $SU(p)$ and $SU(q)$ root lattices.
It is natural to relate these lattices to
the gauge groups realized on the fixed loci
in $X_7$, and identify $\mu_i$ and $\mu_{i'}$ with
the vectors $\rho_i$ and $\rho_{i'}$
introduced in \S\ref{twisted.sec};
\begin{equation}
\mu_i=\rho_i,\quad
\mu_{i'}=\rho_{i'}.
\label{murho}
\end{equation}
For this identification to be justified,
$(\vec\mu,\vec\mu')$, $P$, and $P'$ should satisfy
the same relation as (\ref{ppselect}).
We can easily confirm this as follows.
For every $I$, the $\beta_{I11}$ integration picks up
terms proportional to
\begin{equation}
z_I^{d-kc_I}.
\end{equation}
Therefore, $P$ and $P'$,
the total numbers of $z$ and $z'$, are given by
\begin{eqnarray}
\label{monquant1}
P&=&\sum_{i=1}^p(d-kc_i)
=pd+\sum_{i'=1}^ql_{i'}\mu_{i'},\\
P'&=&\sum_{i'=1}^q(d-kc_{i'})
=qd+kqm_\bullet-\sum_{i=1}^pl_i\mu_i,
\label{monquant2}
\end{eqnarray}
where $l_I$ are the linking numbers defined by (\ref{linking}).
These equations say that the charge $P$ or $P'$ of a gauge-invariant monopole operator shifts 
from a multiple of $p$ or $q$, respectively, corresponding to its magnetic charges,
whcih was pointed out in \cite{Imamura:2009ur}.  
On the other hand, the selection rules (\ref{monquant1}), (\ref{monquant2}) 
are nothing but the relations (\ref{ppselect}) on the gravity side.

The charge $P_M$ corresponding to the M-momentum is
\begin{equation}
P_M
=m_\bullet-\frac{1}{kq}\sum_{i=1}^pl_i\mu_i
-\frac{1}{kp}\sum_{i'=1}^ql_{i'}\mu_{i'}.
\label{pmmu}
\end{equation}
The right hand side of this equation is
independent of $d$, and a
function of the magnetic charges $m_a$.
Although each of three terms in (\ref{pmmu}) separately depends on the choice of
the reference point,
the sum of them is independent of the choice.

We can obtain the relation
(\ref{pmmu}) in more direct way from (\ref{totindex}).
The reason why $P_M$ is related to $m_a$ is that
the flavor rotation generated by
$P_M$ is gauge equivalent to the shift of the dual photon field $a$
defined by (\ref{dualphoton}).
The gauge invariance of an operator require its
charges associated with these two shifts to be the same.
When the gauge group is $U(1)^r$, the gauge symmetry connecting
these two shifts are parameterized by $\varphi$
defined in (\ref{bpt}).
For $U(N)^r$ gauge group we can define such a parameter $\varphi$
by
\begin{equation}
\partial_\varphi\beta_{ist}=\frac{1}{p},\quad
\partial_\varphi\beta_{i'st}=-\frac{1}{q},\quad
\partial_\varphi\beta_{ast}=0.
\end{equation}
The action of $\partial_\varphi$ on the parameters $\alpha_{as}$ is
\begin{equation}
\partial_\varphi\alpha_{as}=\gamma_a,
\end{equation}
where $\gamma_a$ are constants satisfying
\begin{equation}
\gamma_{L(i)}-\gamma_{R(i)}=\frac{1}{p},\quad
\gamma_{L(i')}-\gamma_{R(i')}=-\frac{1}{q}.
\end{equation}
These conditions determine $\gamma_a$ up to overall shift.
Integration of $\varphi$ leaves only the contribution
of operators which are invariant under
the $\varphi$ gauge transformation,
and reproduces the relation (\ref{pmmu}) as we see below.
Let us perform the integration over $\varphi$ orbit in (\ref{totindex}).
A term proportional to $z^Pz^{P'}$ is accompanied by the factor
$e^{-ikP_M\varphi}$.
The other factor including $\varphi$ is the exponential factor including the
Chern-Simons levels.
It includes
\begin{equation}
\exp\left(i\varphi\sum_{a=1}^rk_am_a\gamma_a\right).
\end{equation}
Therefore, for the term to survive after $\varphi$ integration,
the following relation must hold.
\begin{equation}
P_M=\frac{1}{k}\sum_{a=1}^rk_am_a\gamma_a.
\label{pmkmg}
\end{equation}
This is equivalent to (\ref{pmmu}).
This expression is manifestly independent of the reference point.
Thanks to the constraint (\ref{kmzero}),
(\ref{pmkmg}) is not changed by the overall shift of $\gamma_a$,
and is determined unambiguously.

We can easily show
$(1/k)\sum_ak_a\gamma_a=1$,
and $P_M$ is a weighted average of the magnetic charges.

Now we have established the relation between
quantities $(P_M,\rho_i,\rho_{i'})$ defined on the gravity side
and magnetic charges $m_a$ defined on the gauge theory side.
\begin{equation}
P_M=\frac{1}{k}\sum_{a=1}^rk_a\gamma_am_a,\quad
\rho_I=m_{L(I)}-m_{R(I)}.
\label{chargecorre}
\end{equation}
We use this relation when we compare the indices
in the following sections.

\section{Comparison between graviton index and gauge theory index}\label{compare.sec}
In this section we confirm the complete matching of
the gauge theory index and the multi-particle index.
On the previous section we show that the gauge theory index
is factorized into three parts: neutral, positive, and negative parts.
In the following we first show that the multi-particle index on the
gravity side is also factorized in the same way into three parts,
and then, we confirm the agreement for each factor.

We show the agreement for the neutral part analytically.
Concerning the charged part,
we use computers to compute gauge theory index for
many sectors with different charges,
and we show that the gauge theory
index $I^{(+)}_{\{m_a\}}$ for monopole charges $\{m_a\}$
agrees with the multi-particle index $I^{\rm mp}_{(P_M,\vec\rho,\vec\rho')}$
for the charges $(P_M,\vec\rho,\vec\rho')$ corresponding to the magnetic charges
$\{m_a\}$ through (\ref{chargecorre}).

\subsection{Factorization of multi-particle index}
In the previous section, we show that the gauge theory index
is factorized into three parts.
For the two indices to coincides,
the multi-particle index should also have this property.
Namely, $I^{\rm mp}$ should be factorized as
\begin{equation}
I^{\rm mp}=
I^{{\rm mp}(0)}
I^{{\rm mp}(+)}
I^{{\rm mp}(-)}.
\end{equation}
Let us first confirm this factorization.

The factorization of the multi-particle index
is equivalent to the following decomposition of
the single-particle index
\begin{equation}
I^{\rm sp}=
I^{{\rm sp}(0)}
+I^{{\rm sp}(+)}
+I^{{\rm sp}(-)}.
\label{spdecomp}
\end{equation}

Let us consider a single-particle state with quantum numbers
$(P_M,\vec\rho,\vec\rho')$.
By the relations in (\ref{chargecorre})
we can determine the corresponding magnetic charges $m_a$.
The decomposability
(\ref{spdecomp}) claims that
the magnetic charges $m_a$ determined in this way for
every single-particle state do not include
positive and negative components at the same time.
This is confirmed easily as follows.

For a bulk graviton state, which has vanishing vectors
$\vec\rho=\vec\rho'=0$, all the components of the corresponding
magnetic charge are the same and are given by
\begin{equation}
m_1=\cdots =m_r=P_M,
\end{equation}
and thus they never include both positive and negative charges.
This is also the case for the Cartan part of the twisted sectors.

For an $H_\cS$-charged particle in a twisted sector,
one of $\vec\rho$ and $\vec\rho'$ is non-vanishing.
If the particle corresponds to an $SU(p)$ root vector,
$\rho_i$ has two non-vanishing components,
and one of them is $+1$ and the other is $-1$.
In this case the second relation in (\ref{chargecorre})
means that the minimum and the maximum components
of the magnetic charges $m_a$ differ by only one.
Therefore, the $r$ magnetic charges cannot include both
positive and negative charges.

We can always classify single-particle states
into neutral, positive, and negative parts
according to the magnetic charges,
and correspondingly, we can decompose the
single-particle index into the three parts
as (\ref{spdecomp}).

\subsection{Neutral part}
For the neutral part, we can analytically
prove the relation $I^{(0)}=I^{{\rm mp}(0)}$ as we demonstrate below.

On the gravity side,
the neutral part of the multi-particle index $I^{{\rm mp}(0)}$
is given by
\begin{equation}
I^{{\rm mp}(0)}(x,z,z')=\exp\sum_{n=1}^\infty\frac{1}{n}
I_{(0,\vec0,\vec0)}^{\rm sp}(x^n,z^n,z'^n)
\label{imp0}
\end{equation}
where the single-particle index for $(P_M,\vec\rho,\vec\rho')=(0,\vec0,\vec0)$
is (See (\ref{ispppp}).)
\begin{eqnarray}
I_{(0,\vec0,\vec0)}^{\rm sp}(x,z,z')
&=&\sum_{a=-\infty}^\infty I^{\rm grav}_{pa,qa}(x)z^{pa}z'^{qa}
+(p-1)I^{\rm vec}_0(x)
+(q-1)I^{\rm vec}_0(x)
\nonumber\\
&=&
\frac{x^{p+q}z^pz'^q}{1-x^{p+q}z^pz'^q}
+\frac{x^{p+q}z^{-p}z'^{-q}}{1-x^{p+q}z^{-p}z'^{-q}}
+(p+q)\frac{x^2}{1-x^4}.
\label{isp0}
\end{eqnarray}
The corresponding multi-particle index defined by (\ref{imp0}) is
\begin{equation}
I^{{\rm mp}(0)}=\prod_{i=1}^\infty\frac{(1+x^2)^{i(p+q)}}{
(1-(x^{p+q}z^pz'^q)^i)
(1-(x^{p+q}z^{-p}z'^{-q})^i)},
\end{equation}
where we used Euler's partition identity to obtain this expression.

On the gauge theory side,
the corresponding index (\ref{izero}) is
\begin{equation}
I^{(0)}(x,z,z')=\prod_{n=1}^\infty\frac{1}{\det M(x^n,z^n,z'^n)},
\label{i0z}
\end{equation}
where $M$ is the matrix defined in (\ref{mdef}).
We can easily compute the determinant
by rewriting the matrix $M$ as
\begin{equation}
M=\frac{1}{1+x^2}(1-xA)(1-xA^{-1})
\end{equation}
with the matrix
\begin{equation}
A(z,z')=\left(
\begin{array}{cccccc}
\ddots \\
& 0 & z_{I-1} \\
&& 0 & z_I \\
&&& 0 & z_{I+1} \\
&&&& 0 &  \\
z_{r}&&&&& \ddots \\
\end{array}
\right).
\end{equation}
The determinant
\begin{equation}
\frac{1}{\det M}=\frac{(1+x^2)^{p+q}}{
(1-x^{p+q}z^pz'^q)
(1-x^{p+q}z^{-p}z'^{-q})}
\end{equation}
does not depend on the
order of the untwisted and twisted hypermultiplets
in the quiver diagram.
On substituting this into (\ref{i0z}),
we see that the neutral part of the gauge theory index
actually coincides with the corresponding part of the graviton index;
\be
I^{(0)}(x,z,z')=I^{{\rm mp}(0)}(x,z,z').
\ee
This result is consistent with the result in \cite{Choi:2008za}.
If we set $z=z'=1$ and $p=q=M$, we reproduce the
index (\ref{n4sfindex}) with $y_1=1$ substituted.

\subsection{Charged part}
Next, let us confirm the agreement of the charged part:
\begin{equation}
I^{(\pm)}(x,z,z')=
I^{{\rm mp}(\pm)}(x,z,z').
\end{equation}
We can easily show the following relations between positive and
the negative parts:
\begin{equation}
I^{(+)}(x,z,z')=I^{(-)}(x,z^{-1},z'^{-1}),\quad
I^{{\rm mp}(+)}(x,z,z')=I^{{\rm mp}(-)}(x,z^{-1},z'^{-1}).
\end{equation}
Therefore, it is enough to show the relation for the positive part of
the indices:
\begin{equation}
I_{\{m_a\}_+}^{(+)}(x,z,z') =I^{{\rm mp}(+)}_{(P_M,{\vec \rho},{\vec \rho'})}(x,z,z'),
\label{final+}
\end{equation}
for $\{m_a\}_+$ and $(P_M,{\vec \rho},{\vec \rho'})$
related by (\ref{chargecorre}).

Unfortunately, we have not succeeded in proving
(\ref{final+}) analytically.
In the following,
we consider three examples of ${\cal N}=4$ Chern-Simons theories
specified by $\{s_I\}=\{0,0,1\}$, $\{0,0,1,1\}$,
and $\{0,1,0,1\}$
\footnote{%
When we describe a set of numbers $x_a$ assigned to vertices
in the quiver diagram,
we choose a reference vertex $a=\bullet$, which is also used for the
definition of the linking numbers,
and represent $\{x_a\}$ as the vector
$\{x_\bullet,x_{R^2(\bullet)},\ldots,x_{L^2(\bullet)}\}$,
where $R^2(\bullet)\equiv R(R(\bullet))$, and $L^n(\bullet)$ and $R^n(\bullet)$
are similarly defined.
For a set of numbers $y_I$ assigned to edges,
we represent them as
$\{y_{R(\bullet)},y_{R^3(\bullet)},\ldots,y_{L(\bullet)}\}$.}.
(The simplest case with $\{s_I\}=\{0,1\}$ (ABJM model)
has already been investigated in \cite{Kim:2009wb}.)
For each theory
we compute $I^{(+)}_{\{m_a\}}$ numerically
for many sectors specified by
the charges, and confirm the agreement with $I^{{\rm mp}(+)}_{(P_M,\vec\rho,\vec\rho')}$.

\subsubsection{UUT theory}
In this section, we consider the theory defined by
\begin{equation}
\{s_I\}=\{0,0,1\}.
\end{equation}
The background geometry of this theory
is $(\CC^2/\ZZ_2\times \CC^2)/\ZZ_k$.
The internal space $X_7$ includes a $\ZZ_2$-fixed singular locus,
and there exists one two-cycle at the locus $\cSU$.
The vectors $\vec\rho=\{\rho_i\}$ and $\vec\rho'=\{\rho_{i'}\}$
are parameterized by a single winding number $\rho \in\ZZ$
as
\begin{equation}
\vec\rho=\{\rho_1,\rho_2\}=\{-\rho,\rho\},\quad
\vec\rho'=\{\rho_3\}=\{0\}.
\label{vecrhouut}
\end{equation}
We introduce chemical potential $t$ for the
charge $\rho$.
This is related to the potentials $t_I$ introduced in \S\ref{grav.sec}
by $t=t_2/t_1$.
By the relations in (\ref{chargecorre}),
the magnetic charges are determined as
\begin{equation}
\{m_1,m_2,m_3\}=\{P_M,P_M+\rho,P_M\}.
\end{equation}
The Wilson lines $\eta_I$ vanish up to integers,
and this is consistent with the fact that there is no three-cycles
in the dual geometry.
The quantization rules (\ref{monquant1}) and (\ref{monquant2}) for the charges $P$ and $P'$ are
\begin{equation}
P=2a,\quad
P'=a+kP_M,\quad
a, P_M\in\ZZ.
\end{equation}

The positive part of the single-particle index is defined by
$m_a\geq0$ and $\{m_1,m_2,m_3\}\neq\{0,0,0\}$.
These conditions mean
\begin{equation}
P_M\geq 0,\quad
P_M+\rho\geq 0,\quad
(P_M,\rho)\neq(0,0).
\label{uutbounds}
\end{equation}
For every pair of charges $(P_M,\rho)$ satisfying
(\ref{uutbounds})
we would like to confirm
\begin{equation}
I^{(+)}_{\{P_M,P_M+\rho,P_M\}}(x,z,z')
=I^{{\rm mp}(+)}_{(P_M,\rho)}(x,z,z').
\end{equation}
Single-particle states exist only for $|\rho|\leq1$.
Eq. (\ref{ispppp}) gives
\begin{eqnarray}
I^{\rm sp}_{(P_M,0)}
&=&\sum_{a=-\infty}^\infty I^{\rm grav}_{2a,a+kP_M}(x)z^{2a}z'^{kP_M+a}
+I^{\rm vec}_{kP_M}(x)z'^{kP_M},\\
I^{{\rm sp}}_{(P_M,\pm1)}
&=&I^{\rm vec}_{kP_M}(x)z'^{kP_M}.
\end{eqnarray}

It is relatively easy to compute
indices when one of two bounds in (\ref{uutbounds})
is saturated.
Let us first consider $P_M=0$ case.
In this case, we should confirm
\begin{equation}
I^{(+)}_{\{0,\rho,0\}}(x,z,z')
=I^{{\rm mp}(+)}_{(0,\rho)}(x,z,z').
\label{0w0}
\end{equation}
Because the single-particle index depends on the level $k$ only through
the combination $P_Mk$, the multi-particle index on the right hand side in
(\ref{0w0}) is independent of $k$.
We can easily see that this is also the case for the
gauge theory index on the left hand side in (\ref{0w0})
from the expression (\ref{ipom}).

The only non-vanishing single particle index for $P_M=0$ contributing to
$I^{{\rm mp}(+)}$ is
\begin{equation}
I^{\rm sp}_{(0,1)}=\frac{x^2}{1-x^4},
\end{equation}
and the multi-particle index with $P_M=0$ is defined by
\begin{equation}
\sum_{\rho=0}^\infty I^{{\rm mp}(+)}_{(0,\rho)}(x,z,z')t^\rho
=\exp\sum_{n=1}^\infty\frac{1}{n}I^{\rm sp}_{(0,1)}(x^n,z^n,z'^n)t^n.
\end{equation}
By using the identity
\begin{equation}
\prod_{i=0}^\infty\frac{1}{1-tx^i}
=\sum_{i=0}^\infty t^i\prod_{j=1}^i\frac{1}{1-x^j},
\label{identity}
\end{equation}
we obtain
\begin{equation}
I^{{\rm mp}(+)}_{(0,\rho)}=\prod_{i=1}^\rho\frac{x^2}{1-x^{4i}}.
\label{mp0w}
\end{equation}
Let us confirm that the gauge theory index agrees with this
for small $\rho$.
For $\rho=1$, we can easily compute the corresponding gauge theory index
by hand, and confirm the agreement.
\begin{equation}
I^{(+)}_{\{0,1,0\}}=I^{(+)}_{(\cdot,\Y(1),\cdot)}
=\frac{x^2}{1-x^4}.
\end{equation}
For $\rho=2$, there are two contribution with different
monopole backgrounds.
\begin{equation}
I^{(+)}_{\{0,2,0\}}
=
I^{(+)}_{\{\cdot,\Y(2),\cdot\}}
+I^{(+)}_{\{\cdot,\Y(1,1),\cdot\}}.
\end{equation}
It is again easy to compute these two contributions by hand.
They are
\begin{equation}
I^{(+)}_{\{\cdot,\Y(2),\cdot\}}=\frac{x^4}{1-x^8},\quad
I^{(+)}_{\{\cdot,\Y(1,1),\cdot\}}=\frac{x^8}{(1-x^4)(1-x^8)},
\end{equation}
and the summation agrees with the multi-particle index
\begin{equation}
I^{(+)}_{\{0,2,0\}}
=\frac{x^2}{1-x^4}\frac{x^2}{1-x^8}
=I^{{\rm mp}(+)}_{(0,2)}.
\end{equation}
As the charge becomes large,
the computation of the gauge theory index
becomes complicated rapidly.
For $\rho\geq3$, we use computers to generate gauge theory index
as series expansion with respect to the variable $x$,
and check the agreement for small $\rho$ up to certain order
of $x$.
The result is as follows.
\begin{eqnarray}
I^{(+)}_{\{0,3,0\}}
&=&
I^{(+)}_{\{\cdot,\Y(3),\cdot\}}
+I^{(+)}_{\{\cdot,\Y(2,1),\cdot\}}
+I^{(+)}_{\{\cdot,\Y(1,1,1),\cdot\}}
\nonumber\\
&=&I^{{\rm mp}(+)}_{(0,3)}+{\cal O}(x^{101}),
\\
I^{(+)}_{\{0,4,0\}}
&=&
I^{(+)}_{\{\cdot,\Y(4),\cdot\}}
+I^{(+)}_{\{\cdot,\Y(3,1),\cdot\}}
+I^{(+)}_{\{\cdot,\Y(2,2),\cdot\}}
+I^{(+)}_{\{\cdot,\Y(2,1,1),\cdot\}}
+I^{(+)}_{\{\cdot,\Y(1,1,1,1),\cdot\}}
\nonumber\\
&=&I^{{\rm mp}(+)}_{(0,4)}+{\cal O}(x^{101}),
\\
I^{(+)}_{\{0,5,0\}}
&=&
I^{(+)}_{\{\cdot,\Y(5),\cdot\}}
+I^{(+)}_{\{\cdot,\Y(4,1),\cdot\}}
+I^{(+)}_{\{\cdot,\Y(3,2),\cdot\}}
+I^{(+)}_{\{\cdot,\Y(3,1,1),\cdot\}}
\nonumber\\&&
+I^{(+)}_{\{\cdot,\Y(2,2,1),\cdot\}}
+I^{(+)}_{\{\cdot,\Y(2,1,1,1),\cdot\}}
+I^{(+)}_{\{\cdot,\Y(1,1,1,1,1),\cdot\}}
\nonumber\\
&=&I^{{\rm mp}(+)}_{(0,5)}+{\cal O}(x^{31}).
\end{eqnarray}
All these results are consistent with
(\ref{mp0w}) up to the order we have computed.

Next let us consider the case with $P_M\geq1$ and $P_M+\rho=0$.
The relation we would like to confirm is
\begin{equation}
I_{\{P_M,0,P_M\}}^{(+)}(x,z,z')=I^{{\rm mp}(+)}_{(P_M,-P_M)}(x,z,z').
\label{m0m}
\end{equation}
The single-particle index contributing to this part is
\begin{equation}
I^{\rm sp}_{(1,-1)}=\frac{x^2(xz')^k}{1-x^4}.
\end{equation}
With the help of the identity (\ref{identity})
we obtain
\begin{equation}
I^{{\rm mp}(+)}_{(P_M,-P_M)}
=\prod_{i=1}^{P_M}\frac{x^{k+2}z'^k}{1-x^{4i}}.
\label{mpmmm}
\end{equation}
We have confirmed the following relations
up to the indicated order of $x$
for $k=1,2,3,4,5$.
\begin{eqnarray}
I^{(+)}_{\{1,0,1\}}
&=&I^{(+)}_{\{\Y(1),\cdot,\Y(1)\}}
\nonumber\\
&=&I^{{\rm mp}(+)}_{(1,-1)}+{\cal O}(x^{101}),
\\
I^{(+)}_{\{2,0,2\}}
&=&I^{(+)}_{\{\Y(2),\cdot,\Y(2)\}}
+I^{(+)}_{\{\Y(1,1),\cdot,\Y(2)\}}
+I^{(+)}_{\{\Y(2),\cdot,\Y(1,1)\}}
+I^{(+)}_{\{\Y(1,1),\cdot,\Y(1,1)\}}
\nonumber\\
&=&I^{{\rm mp}(+)}_{(2,-2)}+{\cal O}(x^{31}),\\
I^{(+)}_{\{3,0,3\}}
&=&
I^{(+)}_{\{\Y(3),\cdot,\Y(3)\}}
+I^{(+)}_{\{\Y(2,1),\cdot,\Y(3)\}}
+I^{(+)}_{\{\Y(1,1,1),\cdot,\Y(3)\}}
+I^{(+)}_{\{\Y(3),\cdot,\Y(2,1)\}}
+I^{(+)}_{\{\Y(2,1),\cdot,\Y(2,1)\}}
\nonumber\\&&
+I^{(+)}_{\{\Y(1,1,1),\cdot,\Y(2,1)\}}
+I^{(+)}_{\{\Y(3),\cdot,\Y(1,1,1)\}}
+I^{(+)}_{\{\Y(2,1),\cdot,\Y(1,1,1)\}}
+I^{(+)}_{\{\Y(1,1,1),\cdot,\Y(1,1,1)\}}
\nonumber\\
&=&I^{{\rm mp}(+)}_{(3,-3)}+{\cal O}(x^{11}).
\end{eqnarray}
All these results are consistent with (\ref{mpmmm}).

Finally, let us consider
a few examples in which all magnetic charges are positive.
For $k=1,2,3,4,5$
we have checked
\begin{eqnarray}
I^{(+)}_{\{1,1,1\}}
&=&I^{(+)}_{\{\Y(1),\Y(1),\Y(1)\}}
\nonumber\\
&=&I^{{\rm mp}(+)}_{(1,0)}+{\cal O}(x^{31}),\\
I^{(+)}_{\{2,1,2\}}
&=&
I^{(+)}_{\{\Y(2),\Y(1),\Y(2)\}}
+I^{(+)}_{\{\Y(1,1),\Y(1),\Y(2)\}}
+I^{(+)}_{\{\Y(2),\Y(1),\Y(1,1)\}}
+I^{(+)}_{\{\Y(1,1),\Y(1),\Y(1,1)\}}
\nonumber\\
&=&I^{{\rm mp}(+)}_{(2,-1)}+{\cal O}(x^{21}),\\
I^{(+)}_{\{2,2,2\}}
&=&
I^{(+)}_{\{\Y(2),\Y(2),\Y(2)\}}
+I^{(+)}_{\{\Y(2),\Y(2),\Y(1,1)\}}
+I^{(+)}_{\{\Y(2),\Y(1,1),\Y(2)\}}
+I^{(+)}_{\{\Y(2),\Y(1,1),\Y(1,1)\}}
\nonumber\\&&
+I^{(+)}_{\{\Y(1,1),\Y(2),\Y(2)\}}
+I^{(+)}_{\{\Y(1,1),\Y(2),\Y(1,1)\}}
+I^{(+)}_{\{\Y(1,1),\Y(1,1),\Y(2)\}}
+I^{(+)}_{\{\Y(1,1),\Y(1,1),\Y(1,1)\}}
\nonumber\\
&=&I^{{\rm mp}(+)}_{(2,0)}+{\cal O}(x^{11}).
\end{eqnarray}
where
\begin{eqnarray}
I^{{\rm mp}(+)}_{(1,0)}
&=&I^{\rm sp}_{(1,0)}
+I^{\rm sp}_{(0,1)}I^{\rm sp}_{(1,-1)},\\
I^{{\rm mp}(+)}_{(2,-1)}
&=&I^{\rm sp}_{(2,-1)}+I^{\rm sp}_{(1,0)}I^{\rm sp}_{(1,-1)}
+I^{\rm sp}_{(0,1)}
\left(\frac{1}{2}(I^{\rm sp}_{(1,-1)})^2+\frac{1}{2}I^{\rm sp}_{(1,-1)}(\cdot^2)\right),\\
I^{{\rm mp}(+)}_{(2,0)}
&=&
I^{\rm sp}_{(2,0)}
+I^{\rm sp}_{(2,-1)}I^{\rm sp}_{(0,1)}
+I^{\rm sp}_{(1,1)}I^{\rm sp}_{(1,-1)}
\nonumber\\&&
+\left(\frac{1}{2}(I^{\rm sp}_{(1,-1)})^2+\frac{1}{2}I^{\rm sp}_{(1,-1)}(\cdot^2)\right)
\left(\frac{1}{2}(I^{\rm sp}_{(0,1)})^2+\frac{1}{2}I^{\rm sp}_{(0,1)}(\cdot^2)\right)
\nonumber\\&&
+\frac{1}{2}(I^{\rm sp}_{(1,0)})^2
+\frac{1}{2}I^{\rm sp}_{(1,0)}(\cdot^2)
+I^{\rm sp}_{(1,0)}I^{\rm sp}_{(1,-1)}I^{\rm sp}_{(0,1)}.
\end{eqnarray}

\subsubsection{UUTT theory}
Next, let us consider the cases with $p=2$ and $q=2$.
There are two cases with $\{s_I\}=\{0,0,1,1\}$ and $\{s_I\}=\{0,1,0,1\}$, which we call
UUTT and UTUT theories, respectively.
These are simplest examples that are distinguished by the order of
two kinds of hypermultiplets in the quiver diagrams.

We first consider UUTT theory with $\{s_I\}=\{0,0,1,1\}$.
The inking numbers are
\begin{equation}
\vec l=\{l_1,l_2\}=\{2k,2k\},\quad
\vec l'=\{l_3,l_4\}=\{-2k,-2k\},
\end{equation}
and the Wilson line parameters $\eta_I$ vanishes up to integers.
On the gravity side, we have two $A_1$ type singular loci.
We parameterize the vectors $\vec\rho$ and $\vec\rho'$ by
two integers $\rho$ and $\rho'$ as
\begin{equation}
\vec\rho=\{\rho_1,\rho_2\}=\{-\rho,\rho\},\quad
\vec\rho'=\{\rho_3,\rho_4\}=\{-\rho',\rho'\},
\end{equation}
We introduce chemical potentials $t$ and $t'$ for the
charges $\rho$ and $\rho'$, respectively.
These are related to the potentials $t_I$ introduced in \S\ref{grav.sec}
by $t=t_2/t_1$ and $t'=t_4/t_3$.
Eq. (\ref{chargecorre}) gives
\begin{equation}
\{m_1,m_2,m_3,m_4\}=\{P_M,P_M+\rho,P_M,P_M+\rho'\}.
\end{equation}
The positive part is defined by
\be
m_a\geq0,\quad
\{m_1,m_2,m_3,m_4\}\neq\{0,0,0,0\},
\label{positivem}
\ee
and these are equivalent to
\begin{equation}
P_M\geq 0,\quad
P_M+\rho\geq 0,\quad
P_M+\rho'\geq 0,\quad
(P_M,\rho,\rho')\neq(0,0,0).
\label{drrcond}
\end{equation}
We would like to show
\begin{equation}
I^{(+)}_{\{P_M,P_M+\rho,P_M,P_M+\rho'\}}(x,z,z')
=I^{{\rm mp}(+)}_{(P_M,\rho,\rho')}(x,z,z'),
\end{equation}
for every set of charges $(P_M,\rho,\rho')$ satisfying (\ref{drrcond}).
Eq (\ref{ispppp}) gives the single-particle index
\bea
I^{\rm sp}_{(P_M,0,0)}(x,z,z')
&=&\sum_{a=-\infty}^\infty I^{\rm grav}_{2a,2(kP_M+a)}(x)z^{2a}z'^{2(kP_M+a)}
\nonumber\\&&
+I^{\rm vec}_{2kP_M}(x)(z'^{2kP_M}+z^{-2kP_M}),\\
I^{\rm sp}_{(P_M,-1,0)}(x,z,z')
&=&I^{\rm vec}_{2kP_M}(x)z'^{2kP_M},\\
I^{\rm sp}_{(P_M,0,-1)}(x,z,z')
&=&I^{\rm vec}_{-2kP_M}(x)z^{-2kP_M},\\
I^{\rm sp}_{(P_M,1,0)}(x,z,z')
&=&I^{\rm vec}_{2kP_M}(x)z'^{2kP_M},\\
I^{\rm sp}_{(P_M,0,1)}(x,z,z')
&=&I^{\rm vec}_{-2kP_M}(x)z^{-2kP_M}.
\eea

When some of the inequalities
in (\ref{drrcond})
are saturated, the computation of
$I^{{\rm mp}(+)}$ and $I^{(+)}$ are relatively easy,
and we first consider such cases.
The last condition in (\ref{drrcond}) means that the first three
inequalities are not saturated at the same time.
If $P_M=0$,
only single-particle states saturating the same inequality
can contribute to the multi-particle index.
There are only two such single-particle charges,
$(P_M,\rho,\rho')=(0,1,0)$ and $(0,0,1)$,
and thus the multi-particle index is given by
\be
\sum_{\rho=0}^\infty
\sum_{\rho'=0}^\infty
 I^{{\rm mp}(+)}_{(0,\rho,\rho')} t^\rho t'^{\rho'}= 
\exp\left(\sum_{n\geq1} {1\over n} \biggl[I^{\rm sp}_{(0,1,0)}(x^n,z^n,z'^n) t^n 
+I^{\rm sp}_{(0,0,1)}(x^n,z^n,z'^n) t'^n\biggr]\right)
\ee
By using the identity (\ref{identity}), we obtain
\begin{equation}
I^{{\rm mp}(+)}_{(0,\rho,\rho')}=
\left(\prod_{i=1}^\rho\frac{x^2}{1-x^{4i}}\right) 
\left(\prod_{i'=1}^{\rho'}\frac{x^2}{1-x^{4i'}}\right).
\label{mp0rhorhop}
\end{equation}
We can easily show that for $I^{(+)}_{\{0,\rho,0,\rho'\}}$
the integrals in (\ref{ipom}) are factorized into two parts,
and the relation
\begin{equation}
I^{(+)}_{\{0,\rho,0,\rho'\}}=I^{(+)}_{\{0,\rho,0,0\}}I^{(+)}_{\{0,0,0,\rho'\}}
\label{factorization0x0x}
\end{equation}
holds.
In general, if the cyclic sequence of the magnetic charges splits
into several parts
by vanishing components,
the integrals in (\ref{ipom}) are factorized,
and we obtain a relation like (\ref{factorization0x0x}).
Furthermore, each of two factors in (\ref{factorization0x0x})
is the same as the index $I^{(+)}_{\{0,\rho,0\}}$ for the UUT theory.
By using the results in the last subsection,
we can confirm $I^{{\rm mp}(+)}_{(0,\rho,\rho')}=I^{(+)}_{\{0,\rho,0,\rho'\}}$.

Next, let us consider the case in which $P_M\geq1$ and
the second or the third bounds in (\ref{drrcond}) are saturated.
Namely, $P_M+\rho=0$ or $P_M+\rho'=0$.
Because there is no one particle state saturating both the bounds,
the multi-particle index for such charges vanishes;
$I^{{\rm mp}(+)}_{(P_M,-P_M,-P_M)}=0$.
On the gauge theory side, we can show $I^{(+)}_{\{P_M,0,P_M,0\}}=0$
by using the factorization
$I^{(+)}_{\{P_M,0,P_M,0\}}
=I^{(+)}_{\{P_M,0,0,0\}}I^{(+)}_{\{0,0,P_M,0\}}$,
and applying the selection rules to the two factors.

When only one of
$P_M+\rho=0$ or $P_M+\rho'=0$
in (\ref{drrcond}) is saturated,
only single-particle states with charges $(1,0,-1)$ or $(1,-1,0)$
contribute to the multi-particle index,
and we obtain
\begin{equation}
I^{{\rm mp}(+)}_{(\rho',0,-\rho')}=\left(\prod_{i'=1}^{\rho'}\frac{x^{2}(xz^{-1})^{pk}}{1-x^{4i'}}\right) ,\quad
I^{{\rm mp}(+)}_{(\rho,-\rho,0)}=\left(\prod_{i=1}^{\rho}\frac{x^{2}(xz')^{qk}}{1-x^{4i}}\right).
\end{equation}
These are easily generalized to
\begin{equation}
I^{{\rm mp}(+)}_{(\rho+\rho',-\rho,-\rho')}
=\left(\prod_{i'=1}^{\rho'}\frac{x^{2}(xz^{-1})^{pk}}{1-x^{4i'}}\right) 
\left(\prod_{i=1}^{\rho}\frac{x^{2}(xz')^{qk}}{1-x^{4i}}\right).
\label{mpr+r'r-r'}
\end{equation}
We confirm for $k=1,\ldots,5$ that this index is correctly reproduced as the gauge theory index
for small $\rho$ and $\rho'$ as follows.
\bea
I^{(+)}_{\{1,1,1,0\}}
&=&I^{(+)}_{\{\Y(1),\Y(1),\Y(1),\cdot\}}
\nonumber\\
&=&I_{(1,0,-1)}^{{\rm mp}(+)}+{\cal O}(x^{101}),\\
I^{(+)}_{\{2,2,2,0\}}
&=&I^{(+)}_{\{\Y(2),\Y(2),\Y(2),\cdot\}}
+I^{(+)}_{\{\Y(1,1),\Y(2),\Y(2),\cdot\}}
+I^{(+)}_{\{\Y(2),\Y(1,1),\Y(2),\cdot\}}
+I^{(+)}_{\{\Y(2),\Y(2),\Y(1,1),\cdot\}}\notag\\
&&{}+I^{(+)}_{\{\Y(1,1),\Y(1,1),\Y(2),\cdot\}}
+I^{(+)}_{\{\Y(1,1),\Y(2),\Y(1,1),\cdot\}}
+I^{(+)}_{\{\Y(2),\Y(1,1),\Y(1,1),\cdot\}}
+I^{(+)}_{\{\Y(1,1),\Y(1,1),\Y(1,1),\cdot\}}\notag\\
&=&I_{(2,0,-2)}^{{\rm mp}(+)}+{\cal O}(x^{21}),\\
I^{(+)}_{\{1,0,1,1\}}
&=&I^{(+)}_{\{\Y(1),\cdot,\Y(1),\Y(1)\}}
\nonumber\\
&=&I^{{\rm mp}(+)}_{(1,-1,0)}+{\cal O}(x^{101}),\\
I^{(+)}_{\{2,0,2,2\}}
&=&I^{(+)}_{\{\Y(2),\cdot,\Y(2),\Y(2)\}}
+I^{(+)}_{\{\Y(1,1),\cdot,\Y(2),\Y(2)\}}
+I^{(+)}_{\{\Y(2),\cdot,\Y(1,1),\Y(2)\}}
+I^{(+)}_{\{\Y(2),\cdot,\Y(2),\Y(1,1)\}}\notag\\
&&{}+I^{(+)}_{\{\Y(1,1),\cdot,\Y(1,1),\Y(2)\}}
+I^{(+)}_{\{\Y(1,1),\cdot,\Y(2),\Y(1,1)\}}
+I^{(+)}_{\{\Y(2),\cdot,\Y(1,1),\Y(1,1)\}}
+I^{(+)}_{\{\Y(1,1),\cdot,\Y(1,1),\Y(1,1)\}}\notag\\
&=&I^{{\rm mp}(+)}_{(2,-2,0)}+{\cal O}(x^{21}),\\
I^{(+)}_{\{2,1,2,1\}}
&=&I^{(+)}_{\{\Y(2),\Y(1),\Y(2),\Y(1)\}}
+I^{(+)}_{\{\Y(1,1),\Y(1),\Y(2),\Y(1)\}}
+I^{(+)}_{\{\Y(2),\Y(1),\Y(1,1),\Y(1)\}}
+I^{(+)}_{\{\Y(1,1),\Y(1),\Y(1,1),\Y(1)\}}
\nonumber\\
&=&I_{(2,-1,-1)}^{{\rm mp}(+)} +{\cal O}(x^{21}).
\eea  

Finally, we give more examples without vanishing magnetic charges.
\begin{eqnarray}
I^{(+)}_{\{1,1,1,1\}}
&=&I^{(+)}_{\{\Y(1),\Y(1),\Y(1),\Y(1)\}}
\nonumber\\
&=&
I_{(1,0,0)}^{{\rm mp}(+)}
+{\cal O}(x^{101}),\\
I^{(+)}_{\{1,2,1,1\}}
&=&I^{(+)}_{\{\Y(1),\Y(2),\Y(1),\Y(1)\}}+I^{(+)}_{\{\Y(1),\Y(1,1),\Y(2),\Y(1)\}}
\nonumber\\
&=&
I^{{\rm mp}(+)}_{(1,1,0)}
+{\cal O}(x^{31}),\\
I^{(+)}_{\{1,1,1,2\}}
&=&I^{(+)}_{\{\Y(1),\Y(1),\Y(1),\Y(2)\}}+I^{(+)}_{\{\Y(1),\Y(1),\Y(1),\Y(1,1)\}}
\nonumber\\
&=&
I^{{\rm mp}(+)}_{(1,0,1)}
+{\cal O}(x^{31}),\\
I^{(+)}_{\{1,2,1,2\}}
&=&I^{(+)}_{\{\Y(1),\Y(2),\Y(1),\Y(2)\}}+I^{(+)}_{\{\Y(1),\Y(1,1),\Y(1),\Y(2)\}}
+I^{(+)}_{\{\Y(1),\Y(2),\Y(1),\Y(1,1)\}}+I^{(+)}_{\{\Y(1),\Y(1,1),\Y(1),\Y(1,1)\}}
\nonumber\\
&=&
I^{{\rm mp}(+)}_{(1,1,1)}
+{\cal O}(x^{31}),
\end{eqnarray}
where
\begin{eqnarray}
I^{{\rm mp}(+)}_{(1,0,0)}
&=&I^{\rm sp}_{(1,0,0)}
+I^{\rm sp}_{(1,-1,0)}I^{\rm sp}_{(0,1,0)}
+I^{\rm sp}_{(1,0,-1)}I^{\rm sp}_{(0,0,1)},\\
I^{{\rm mp}(+)}_{(1,1,0)}
&=&I^{\rm sp}_{(1,1,0)}
+I^{\rm sp}_{(1,0,0)}I^{\rm sp}_{(0,1,0)}
+I^{\rm sp}_{(1,0,-1)}I^{\rm sp}_{(0,0,1)}I^{\rm sp}_{(0,0,1)}\notag\\
&&+\left(\frac{1}{2}(I^{\rm sp}_{(0,1,0)})^2+\frac{1}{2}I^{\rm sp}_{(0,1,0)}(\cdot^2)\right)
I^{\rm sp}_{(1,0,-1)},\\
I^{{\rm mp}(+)}_{(1,0,1)}
&=&I^{\rm sp}_{(1,0,1)}
+I^{\rm sp}_{(1,0,0)}I^{\rm sp}_{(0,0,1)}
+I^{\rm sp}_{(1,-1,0)}I^{\rm sp}_{(0,1,0)}I^{\rm sp}_{(0,1,0)}\notag\\
&&+\left(\frac{1}{2}(I^{\rm sp}_{(0,0,1)})^2+\frac{1}{2}I^{\rm sp}_{(0,0,1)}(\cdot^2)\right)
I^{\rm sp}_{(1,-1,1)},\\
I^{{\rm mp}(+)}_{(1,1,1)}
&=&I^{\rm sp}_{(1,0,0)} I^{\rm sp}_{(0,1,0)}I^{\rm sp}_{(0,0,1)}
+I^{\rm sp}_{(1,0,1)}I^{\rm sp}_{(0,1,0)}+I^{\rm sp}_{(1,1,0)}I^{\rm sp}_{(0,0,1)}
\notag\\
&&+\left(\frac{1}{2}(I^{\rm sp}_{(0,1,0)})^2+\frac{1}{2}I^{\rm sp}_{(0,1,0)}(\cdot^2)\right)
I^{\rm sp}_{(0,0,1)}I^{\rm sp}_{(1,-1,0)}\notag\\
&&+\left(\frac{1}{2}(I^{\rm sp}_{(0,0,1)})^2+\frac{1}{2}I^{\rm sp}_{(0,0,1)}(\cdot^2)\right)
I^{\rm sp}_{(0,1,0)}I^{\rm sp}_{(1,0,-1)}.
\end{eqnarray}

\subsubsection{UTUT theory}
Now we move to the UTUT theory
with $\{s_I\}=\{0,1,0,1\}$.
The linking numbers for this theory are
\begin{equation}
\vec l=\{l_1,l_3\}=\{2k,k\},\quad
\vec l'=\{l_2,l_4\}=\{-k,-2k\},
\end{equation}
and the Wilson line parameters are given by
\begin{equation}
\vec\eta
=\{\eta_1,\eta_3\}
=\{0,\frac{1}{2}\},\quad
\vec\eta'
=\{\eta_2,\eta_4\}
=\{\frac{1}{2},0\}.
\end{equation}
This theory is the simplest example with the non-trivial Wilson
lines on the singular loci.
We parameterize $\vec\rho$ and $\vec\rho'$ by two integers $\rho$ and $\rho'$ as
\begin{equation}
\vec\rho=\{\rho_1,\rho_3\}=\{-\rho,\rho\},\quad
\vec\rho'=\{\rho_2,\rho_4\}=\{-\rho',\rho'\}.
\end{equation}
We introduce chemical potentials $t$ and $t'$ for the
charges $\rho$ and $\rho'$, respectively.
These are related to the potentials $t_I$ introduced in \S\ref{grav.sec}
by $t=t_3/t_1$ and $t'=t_4/t_2$.
Then the magnetic charges are given by
\begin{eqnarray}
\{m_a\}
&=&\{P_M-\frac{\rho+\rho'}{2},P_M+\frac{\rho-\rho'}{2},
P_M+\frac{\rho+\rho'}{2},P_M-\frac{\rho-\rho'}{2}\}
\nonumber\\
&=&\{m_\bullet,m_\bullet+\rho,m_\bullet+\rho+\rho',m_\bullet+\rho'\},
\end{eqnarray}
where $m_\bullet$ is the magnetic charge for the reference vertex,
and is related to $P_M$ by
\bea
P_M= m_\bullet+\frac{1}{2}(\rho+\rho').
\eea

The relation we would like to confirm is
\begin{equation}
I^{(+)}_{\{P_M-\frac{\rho+\rho'}{2},
P_M+\frac{\rho-\rho'}{2},
P_M+\frac{\rho+\rho'}{2},
P_M-\frac{\rho-\rho'}{2}\}}(x,z,z')
=I^{{\rm mp}(+)}_{(P_M,\rho,\rho')}(x,z,z').
\end{equation}
The positive part of the single particle index
is defined by
\be
m_a\geq0,\quad\{m_1,m_2,m_3,m_4\}\neq\{0,0,0,0\},
\ee
and these are equivalent to
\begin{equation}
m_\bullet\geq0,\quad
m_\bullet+\rho\geq0,\quad
m_\bullet+\rho'\geq 0,\quad
(m_\bullet,\rho,\rho')\neq(0,0,0).
\label{ututbound}
\end{equation}

The single particle index is given by
\bea
I^{\rm sp}_{(m_\bullet,0,0)}(x,z,z')
&=&\sum_{a=-\infty}^\infty I^{\rm grav}_{2a,2(km_\bullet+a)}(x)z^{2a}z'^{2(km_\bullet+a)}
\nonumber\\&&
+I^{\rm vec}_{2km_\bullet}(x)(z'^{2km_\bullet}+z^{-2km_\bullet}),\\
I^{\rm sp}_{(m_\bullet-\frac{1}{2},-1,0)}(x,z,z')
&=&I^{\rm vec}_{k(2m_\bullet-1)}(x)z'^{k(2m_\bullet-1)},\\
I^{\rm sp}_{(m_\bullet-\frac{1}{2},0,-1)}(x,z,z')
&=&I^{\rm vec}_{-k(2m_\bullet-1)}(x)z^{-k(2m_\bullet-1)},\\
I^{\rm sp}_{(m_\bullet+\frac{1}{2},1,0)}(x,z,z')
&=&I^{\rm vec}_{k(2m_\bullet+1)}(x)z'^{k(2m_\bullet+1)},\\
I^{\rm sp}_{(m_\bullet+\frac{1}{2},0,1)}(x,z,z')
&=&I^{\rm vec}_{-k(2m_\bullet+1)}(x)z^{-k(2m_\bullet+1)}.
\eea

As we did in the UUTT theory,
let us first consider the cases in which some of
the magnetic charges $m_a$ vanish.
Because all four vertices in the quiver diagram are
on an equal footing,
we can assume $m_\bullet=m_1=0$ without loosing generality.
This means that the first bound in (\ref{ututbound}) is saturated,
and
on the gravity side
only single-particle states with
$(m_\bullet,\rho,\rho')=(0,1,0)$ or
$(0,0,1)$ can contribute to the index.
The multi-particle index in this case is determined with the relation
\bea
&&\sum_{\rho,\rho'} I^{{\rm mp}(+)}_{(\frac{\rho+\rho'}{2},\rho,\rho')}(x,z,z')
t^\rho t'^{\rho'}\notag\\
&=&\exp\left(\sum_{n\geq1} {1\over n} \left[
I^{\rm sp}_{(1/2,1,0)}(x^n,z^n,z'^n)t^n 
+I^{\rm sp}_{(1/2,0,1)}(x^n,z^n,z'^n)t'^n\right]\right).
\eea
By using the identity (\ref{identity}), we obtain
\begin{equation}
I^{{\rm mp}(+)}_{(\frac{\rho+\rho'}{2},\rho,\rho')}=
\left(\prod_{i=1}^\rho\frac{x^2(x z')^k}{1-x^{4i}}\right) 
\left(\prod_{i'=1}^{\rho'}\frac{x^2 (x z^{-1})^k}{1-x^{4i'}}\right).
\label{mp0rhorho2}
\end{equation}
For $k=1,\ldots,5$ and
small $\rho$ and $\rho'$,
we confirmed
that this multi-particle index is reproduced as the gauge theory index
\bea
I^{(+)}_{\{0,0,1,1\}}
&=&I^{(+)}_{\{\cdot,\cdot,\Y(1),\Y(1)\}}
\nonumber\\
&=&I^{{\rm mp}(+)}_{(1/2,0,1)}+{\cal O}(x^{101}),\\
I^{(+)}_{\{0,0,2,2\}}
&=&I^{(+)}_{\{\cdot,\cdot,\Y(2),\Y(2)\}}
+I^{(+)}_{\{\cdot,\cdot,\Y(2),\Y(1,1)\}}
+I^{(+)}_{\{\cdot,\cdot,\Y(1,1),\Y(2)\}}
+I^{(+)}_{\{\cdot,\cdot,\Y(1,1),\Y(1,1)\}}\notag\\
&=&I^{{\rm mp}(+)}_{(1,0,2)}+{\cal O}(x^{31}),\\
I^{(+)}_{\{0,0,3,3\}}
&=&
+I^{(+)}_{\{\cdot,\cdot,\Y(3),\Y(3)\}}
+I^{(+)}_{\{\cdot,\cdot,\Y(3),\Y(2,1)\}}
+I^{(+)}_{\{\cdot,\cdot,\Y(3),\Y(1,1,1)\}}
+I^{(+)}_{\{\cdot,\cdot,\Y(2,1),\Y(3)\}}
+I^{(+)}_{\{\cdot,\cdot,\Y(2,1),\Y(2,1)\}}
\notag\\&&
+I^{(+)}_{\{\cdot,\cdot,\Y(2,1),\Y(1,1,1)\}}
+I^{(+)}_{\{\cdot,\cdot,\Y(1,1,1),\Y(3)\}}
+I^{(+)}_{\{\cdot,\cdot,\Y(1,1,1),\Y(2,1)\}}
+I^{(+)}_{\{\cdot,\cdot,\Y(1,1,1),\Y(1,1,1)\}}
\notag\\
&=&I_{(3/2,0,3)}^{{\rm mp}(+)}+{\cal O}(x^{11}),\\
I^{(+)}_{\{0,1,1,0\}}
&=&I^{(+)}_{\{\cdot,\Y(1),\Y(1),\cdot\}}
\nonumber\\
&=&I^{{\rm mp}(+)}_{(1/2,1,0)}+{\cal O}(x^{101}),\\
I^{(+)}_{\{0,2,2,0\}}
&=&I^{(+)}_{\{\cdot,\Y(2),\Y(2),\cdot\}}
+I^{(+)}_{\{\cdot,\Y(2),\Y(1,1),\cdot\}}
+I^{(+)}_{\{\cdot,\Y(1,1),\Y(2),\cdot\}}
+I^{(+)}_{\{\cdot,\Y(1,1),\Y(1,1),\cdot\}}\notag\\
&=&I^{{\rm mp}(+)}_{(1,2,0)}+{\cal O}(x^{31}),\\
I^{(+)}_{\{0,3,3,0\}}
&=&
I^{(+)}_{\{\cdot,\Y(3),\Y(3),\cdot\}}
+I^{(+)}_{\{\cdot,\Y(3),\Y(2,1),\cdot\}}
+I^{(+)}_{\{\cdot,\Y(3),\Y(1,1,1),\cdot\}}
+I^{(+)}_{\{\cdot,\Y(2,1),\Y(3),\cdot\}}
+I^{(+)}_{\{\cdot,\Y(2,1),\Y(2,1),\cdot\}}
\notag\\&&
+I^{(+)}_{\{\cdot,\Y(2,1),\Y(1,1,1),\cdot\}}
+I^{(+)}_{\{\cdot,\Y(1,1,1),\Y(3),\cdot\}}
+I^{(+)}_{\{\cdot,\Y(1,1,1),\Y(2,1),\cdot\}}
+I^{(+)}_{\{\cdot,\Y(1,1,1),\Y(1,1,1),\cdot\}}
\notag\\
&=&I^{{\rm mp}(+)}_{(3/2,3,0)}+{\cal O}(x^{11}),\\
I^{(+)}_{\{0,1,2,1\}}
&=&I^{(+)}_{\{\cdot,\Y(1),\Y(2),\Y(1)\}}
+I^{(+)}_{\{\cdot,\Y(1),\Y(1,1),\Y(1)\}}\notag\\
&=&I^{{\rm mp}(+)}_{(1,1,1)}+{\cal O}(x^{41}),\\
I^{(+)}_{\{0,2,3,1\}}
&=&
I^{(+)}_{\{\cdot,\Y(2),\Y(3),\Y(1)\}}
+I^{(+)}_{\{\cdot,\Y(2),\Y(2,1),\Y(1)\}}
+I^{(+)}_{\{\cdot,\Y(2),\Y(1,1,1),\Y(1)\}}
+I^{(+)}_{\{\cdot,\Y(1,1),\Y(3),\Y(1)\}}
+I^{(+)}_{\{\cdot,\Y(1,1),\Y(2,1),\Y(1)\}}
+I^{(+)}_{\{\cdot,\Y(1,1),\Y(1,1,1),\Y(1)\}}
\notag\\
&=&I^{{\rm mp}(+)}_{(3/2,2,1)}+{\cal O}(x^{21}),\\
I^{(+)}_{\{0,1,3,2\}}
&=&
I^{(+)}_{\{\cdot,\Y(1),\Y(3),\Y(2)\}}
+I^{(+)}_{\{\cdot,\Y(1),\Y(3),\Y(1,1)\}}
+I^{(+)}_{\{\cdot,\Y(1),\Y(2,1),\Y(2)\}}
+I^{(+)}_{\{\cdot,\Y(1),\Y(2,1),\Y(1,1)\}}
+I^{(+)}_{\{\cdot,\Y(1),\Y(1,1,1),\Y(2)\}}
+I^{(+)}_{\{\cdot,\Y(1),\Y(1,1,1),\Y(1,1)\}}
\notag\\
&=&I^{{\rm mp}(+)}_{(3/2,1,2)}+{\cal O}(x^{21}).
\eea

We also check in some sectors with
magnetic charges without vanishing components
the gauge theory index
correctly reproduces the corresponding multi-particle index
for $k=1,\ldots,5$.
\begin{eqnarray}
I^{(+)}_{\{1,1,1,1\}}
&=&I^{(+)}_{\{\Y(1),\Y(1),\Y(1),\Y(1)\}}
\nonumber\\
&=&
I^{{\rm mp}(+)}_{(1,0,0)}
+{\cal O}(x^{101}),\\
I^{(+)}_{\{1,2,2,1\}}
&=&I^{(+)}_{\{\Y(1),\Y(2),\Y(2),\Y(1)\}}+I^{(+)}_{\{\Y(1),\Y(1,1),\Y(2),\Y(1)\}}
+I^{(+)}_{\{\Y(1),\Y(2),\Y(1,1),\Y(1)\}}+I^{(+)}_{\{\Y(1),\Y(1,1),\Y(1,1),\Y(1)\}}
\nonumber\\
&=&
I^{{\rm mp}(+)}_{(3/2,1,0)}
+{\cal O}(x^{31}),\\
I^{(+)}_{\{1,1,2,2\}}
&=&I^{(+)}_{\{\Y(1),\Y(1),\Y(2),\Y(2)\}}+I^{(+)}_{\{\Y(1),\Y(1),\Y(1,1),\Y(2)\}}
+I^{(+)}_{\{\Y(1),\Y(1),\Y(2),\Y(1,1),\Y(1)\}}+I^{(+)}_{\{\Y(1),\Y(1),\Y(1,1),\Y(1,1)\}}
\nonumber\\
&=&
I^{{\rm mp}(+)}_{(3/2,0,1)}
+{\cal O}(x^{31}),
\end{eqnarray}
where
\bea
I^{{\rm mp}(+)}_{(1,0,0)}
&=&I^{\rm sp}_{(1,0,0)}
+I^{\rm sp}_{(1/2,-1,0)}I^{\rm sp}_{(1/2,1,0)}
+I^{\rm sp}_{(1/2,0,-1)}I^{\rm sp}_{(1/2,0,1)},\\
I^{{\rm mp}(+)}_{(3/2,1,0)}
&=&I^{\rm sp}_{(3/2,1,0)}
+I^{\rm sp}_{(1,0,0)}I^{\rm sp}_{(1/2,1,0)}
+I^{\rm sp}_{(1/2,1,0)}I^{\rm sp}_{(1/2,0,1)}I^{\rm sp}_{(1/2,0,-1)}\notag\\
&&+\left(\frac{1}{2}(I^{\rm sp}_{(1/2,1,0)})^2+\frac{1}{2}I^{\rm sp}_{(1/2,1,0)}(\cdot^2)\right)
I^{\rm sp}_{(1/2,-1,0)},\\
I^{{\rm mp}(+)}_{(3/2,0,1)}
&=&I^{\rm sp}_{(3/2,0,1)}
+I^{\rm sp}_{(1,0,0)}I^{\rm sp}_{(1/2,0,1)}
+I^{\rm sp}_{(1/2,0,1)}I^{\rm sp}_{(1/2,1,0)}I^{\rm sp}_{(1/2,-1,0)}\notag\\
&&+\left(\frac{1}{2}(I^{\rm sp}_{(1/2,0,1)})^2+\frac{1}{2}I^{\rm sp}_{(1/2,0,1)}(\cdot^2)\right)
I^{\rm sp}_{(1/2,0,-1)}.
\eea

\section{Conclusions}\label{conc.sec}
In this paper
we have computed the
index (\ref{igauge}) for ${\cal N}=4$ Chern-Simons theories
with taking account of monopole contribution,
and compared it with the multi-particle index
for M-theory in the background
AdS$_4\times X_7$, where $X_7=({\bf S}^7/(\ZZ_p \times \ZZ_q))/\ZZ_k$.

When we calculated the gauge theory index $I^{\rm gauge}$,
we took the large $N$ limit with
the
Chern-Simons couplings $k_a$ fixed.

On the gravity side, the internal space $X_7$ includes
fixed loci on which
$G_\cS=SU(p)\times SU(q)$ vector multiplets live.
We conjectured the single-particle index
$I^{\rm vec}$ in (\ref{vecons3})
for a vector multiplet in AdS$_4\times{\bf S}^3$,
and derived the contribution of the twisted sectors
by the orbifold projection from $I^{\rm vec}$.
We also derived the bulk sector index as the orbifold projection
of the known graviton index for AdS$_4\times{\bf S}^7$.
We combined them to obtain the multi-particle index $I^{\rm mp}$.

Both the gauge theory index $I^{\rm gauge}$ and the graviton index $I^{\rm mp}$
are factorized into three parts:
neutral, positive, and negative parts.
We analytically proved the agreement of the neutral part of these indices.
The agreement of the negative part follows from that of the positive part.
To compute the positive part of the gauge theory index
we used numerical methods.
We considered three ${\cal N}=4$ Chern-Simons theories
with gauge group $U(N)^3$ or $U(N)^4$ as examples,
and for each theory
we numerically computed the gauge theory index
for various sectors specified by the magnetic charges $\{m_a\}$.
The magnetic charges $\{m_a\}$ are related to
the M-momentum $P_M$ and the $H_\cS$-charges $\rho_I$, where
$H_\cS=U(1)^{r-2}$ is the Cartan subgroup of $G_\cS$.
We conjectured the one-to-one correspondence (\ref{chargecorre}) between
$\{m_a\}$ and $(P_M,\rho_I)$,
and comfirmed that the gauge theory index for a sector
with magnetic charges $\{m_a\}$ agrees with the multi-particle
index for the sector with corresponding charges $(P_M,\rho_I)$.

The $H_\cS$ charges $\rho_I$ can be regarded as the winding numbers of
M2-branes on non-trivial two-cycles, and these results strongly suggest
that (a part of) monopole
operators correspond to wrapped M2-branes.

We also confirmed that the relation between the fivebrane linking numbers
and the $H_\cS$ Wilson lines on the singular loci is reproduced on the
gauge theory side by analyzing the selection rules for the charges
of global symmetry $U(1)\times U(1)'$.

\section*{Acknowledgements}
Y.~I. was supported in part by
Grant-in-Aid for Young Scientists (B) (\#19740122) from the Japan
Ministry of Education, Culture, Sports,
Science and Technology.
S.~Y. was supported by
the Global COE Program ``the Physical Sciences Frontier'', MEXT, Japan.

\end{document}